\documentclass[a4paper,11pt]{article}
\pdfoutput=1 

\usepackage{jheppub} 

\usepackage[T1]{fontenc} 
\usepackage{blindtext}
\usepackage{epstopdf}
\usepackage{graphicx}
\usepackage{epsfig}
\usepackage{dcolumn}  
\usepackage{bm}    
\usepackage{caption}
\usepackage{subcaption}
\usepackage{amssymb} 
\usepackage{tikz}
\usetikzlibrary{arrows}
\usepackage{nccmath}
\usepackage{xcolor}
\usepackage{epstopdf}
\usepackage{graphicx,wrapfig,lipsum}
\usepackage{epsfig}
\usepackage{amsmath,bm}
\usepackage{amsfonts}  
\usepackage{amsmath}  
\usepackage{slashed}  
\usepackage{enumitem}
\usepackage[mathscr]{euscript}
\usepackage{tabu}
\usepackage{epsfig}
\makeatletter
\g@addto@macro\bfseries{\boldmath}
\makeatother

\def\l1{{{1-loop}}}

\def\n1{\Bigg|_{n=1}}

\def\n{{(n)}}

\usepackage[T1]{fontenc} 
\usepackage{tikz}
\usepackage{amsmath,amssymb}
\usepackage{relsize}
\usepackage{latexsym}
\usepackage{leftidx}
\usepackage{xcolor}
\usepackage{diagbox}
\usepackage[T1]{fontenc}
\usepackage{array}
\usepackage{makecell}
\usepackage{csquotes}
\usepackage{tikz}
\usepackage{enumitem}
\usepackage{setspace}
\usepackage{multirow}
\usepackage{amsmath,amssymb}
\usepackage{relsize}
\usepackage{latexsym}
\usepackage{leftidx}
\usepackage{xcolor}
\usepackage{csquotes}
\usepackage{tikz}
\usetikzlibrary{decorations.pathmorphing}
\usetikzlibrary{arrows.meta}
\usepackage{enumitem}
\usetikzlibrary{decorations.markings}
\usetikzlibrary{decorations.pathmorphing}
\usetikzlibrary{decorations.markings}
\usetikzlibrary{decorations.pathmorphing}
\usepackage{pifont}
\usepackage{hyperref}
\usepackage{url}

\usepackage{bookmark}

 \title{\textbf{\textsf{Thermal one-point functions: CFT's with fermions, large $d$ and large spin
 }}}
  \author{Justin R. David, Srijan Kumar}
\affiliation{\vspace{.1cm} Centre for High Energy Physics, \\ Indian Institute of Science,\\
C. V. Raman Avenue, Bangalore 560012, India.}
\emailAdd{justin@iisc.ac.in, srijankumar@iisc.ac.in}

\abstract{
We apply the OPE inversion formula on thermal two-point functions of fermions
to obtain thermal one-point function of fermion bi-linears appearing in the corresponding 
OPE. 
We primarily focus on the OPE channel which contains the stress tensor of the theory. 
We apply our formalism to the mean field theory of fermions and verify that the 
 inversion formula reproduces the spectrum as well as their corresponding thermal one-point functions. 
 We then examine the large $N$ critical Gross-Neveu model in  $d=2k+1$ dimensions  with $k$ even 
 and at finite temperature. We show that stress tensor evaluated from the 
  inversion formula agrees with that evaluated from the partition function at the critical point.
 We demonstrate the expectation values of 3 different classes of higher spin currents are all related to each other by
numerical constants, spin and the thermal mass. 
We evaluate the ratio of the thermal expectation values of 
  higher spin currents at the critical point to the Gaussian fixed point or the Stefan-Boltzmann result,
   both for the large $N$ critical $O(N)$ model and the Gross-Neveu model in odd dimensions. 
   This ratio is always less than one and it approaches unity on increasing the spin with the dimension $d$ held fixed. 
  The ratio however approaches zero when the dimension $d$ is increased with the spin held fixed. 
}

\begin{document} 
\maketitle
\flushbottom

\section{Introduction}

Given a quantum field theory, it is usually important to understand its behaviour at finite temperature. 
That is when one of the directions of the Euclidean theory is taken to be a circle of radius $\beta$, the inverse temperature. 
This question is particularly relevant when the quantum field theory is a conformal field theory. 
Usually critical points of quantum field theories occur at finite temperature. 
Furthermore,  studying conformal field theories which arise in $AdS/CFT$ context implies that one is studying 
properties of $AdS$ black holes. 

It is possible to use the  symmetries of the conformal field theory to constraint conformal field theories on $S^1 \times R^{d-1}$
where the circle $S^1$ is of length $\beta$.  Such a program was initiated in \cite{Iliesiu:2018fao} and pursued in 
\cite{Petkou:2018ynm,Iliesiu:2018zlz,Gobeil:2018fzy,Luo:2022tqy,Benjamin:2023qsc,Marchetto:2023fcw}. 
Under some reasonable assumptions of 
analyticity of the finite temperature 2-point functions  of primary scalar operators, a thermal inversion formula was derived 
\cite{Iliesiu:2018fao}. 
This inversion formula 
allowed one to obtain the thermal one-point functions for all operators which appear in the 
OPE of the given 2 point function. 
The inversion formula was  applied to fermionic 2-point functions in \cite{Petkou:2018ynm}, however the OPE channel studied was 
the scalar channel, that is the spinor indices of the fermionic operators in the 2-point function were contracted. 
This channel for instance does not contain the stress tensor of the theory. 

In this paper we apply the  OPE inversion formula on fermionic 2-point functions and focus on the OPE
channel which contains the stress tensor. 
We will see that there are $2$ classes of operators that exist in this channel, these operators are schematically of the 
form 
\begin{eqnarray}\label{class23}
{\cal O}_+[n, l] &=& \bar \psi \gamma_{\mu_{1}} \partial_{\mu_2} \cdots \partial_{\mu_l}  \partial^{2n} \psi, 
\\ \nonumber
{\cal O}_-[n, l] &=& \bar \psi \gamma^{\mu} \partial_\mu \partial_{\mu_1}
 \partial_{\mu_2} \cdots \partial_{\mu_l}  \partial^{2n} \psi, 
\end{eqnarray}
$\psi$ is a Dirac spinor and these operators are rank $l$ symmetric traceless tensors, $\gamma^\mu$ are the Dirac matrices.
To isolate the one-point functions of operators belonging to each of these classes we need 
to apply the OPE inversion formula to 2  related thermal 2-point functions. 
There is a third class of symmetric traceless tensors schematically of the form 
\begin{equation}\label{class1}
{\cal O }_0[n, l]  = \bar\psi \partial_{\mu_1} \partial_{\mu_2} \cdots \partial_{\mu_l} \partial^{2n} \psi .
\end{equation}
These occur in the scalar channel of the OPE expansion of 2 fermionic operators, to isolate the one-point function of 
these operators we apply the OPE inversion formula to thermal 2-point functions in which the spinor indices are contracted
\cite{Petkou:2018ynm}. 

After presenting the general formalism for  evaluating thermal one-point functions belonging to the classes 
in (\ref{class23}) we apply it to the mean field theory of fermions (MFT).    
We show that 
 the expectation values of operators  obtained by expanding the MFT correlator in the short distance limit 
agrees precisely with that using the OPE inversion formula. 
For $d=2$ this  check is done for all the operators in the class (\ref{class23}), for $d>2$ we perform this 
check for $n=0, 1$.  In MFT, operators in the class (\ref{class1}) do not appear in the OPE. 

We then examine the large $N$  critical Gross-Neveu model in $d = 2k +1$ dimensions. We show that the gap equation 
of the model can be obtained by either demanding operators  $\bar\psi \psi$  or $\bar \psi \gamma^\mu \partial_\mu \psi$ do 
not occur in the spectrum. 
The gap equation has a real solution for the thermal mass  for $k$ even.
 We show that the one-point function of  the operator
${\cal O}_+[0,2]$ precisely agrees with the 
stress tensor obtained from the partition function of the theory. 
The form of expectation value of ${\cal O}_+[0,2]$ appears manifestly different from that of stress tensor 
from the partition function. But   on substituting the value of thermal mass from the gap equation they precisely agree. 
Finally we show that for the large $N$ critical Gross-Neveu model, the expectation values of 
the three classes of operators in (\ref{class23}), (\ref{class1}) all are related by numerical factors, spin and the thermal mass. 
One such relation  we prove using the OPE inversion formula is 
\begin{equation}
a_{{\cal O}_{0}}[0, l] = m_{\rm th} a_{{\cal O}_+}[0, l ] .
\end{equation}
Here $m_{\rm th}$ is the thermal mass and $a_{{\cal O}_0}$ and $a_{{\cal O}_+}$  refer to the thermal expectation values. 
Our analysis shows it is sufficient to work with the operators ${\cal O}_{+}[0, l]$ in this model. 
Finally, we quote here the result for the thermal expectation values of the operators in the class  ${\cal O}_{+}[0, l]$ 
for the critical large $N$ Gross-Neveu model. 
\begin{equation} \label{gnonepint}
a_{{\cal O}_+}[n=0, l] = \frac{l  }{ 2 \pi ^k (  k - \frac{1}{2} )_l} \left( \frac{m_{\rm th}}{2} \right)^{l +k -1} 
\sum_{n=0}^{l +k -1} \frac{ (l +k -n)_{2n}}{ (2 m_{\rm th}) ^n n!} {\rm Li}_{k+n} ( - e^{-m_{\rm th}} ) .
\end{equation}

As we have seen the inversion formula applied to the critical Gross-Neveu as well as the 
$O(N)$ model \cite{Iliesiu:2018fao,Petkou:2018ynm} yields reasonably compact expressions for the one-point function of 
higher spin currents. Since the method can be easily applied to these classes of CFT in any dimensions we can study 
the behaviour of these one-point functions for arbitrary dimensions $d= 2k+1$, and arbitrary spin. 
Motivations to study this include the results from  earlier works related to large $d$ conformal field theories in 
\cite{Fei:2014yja,Fei:2014xta,Stergiou:2015roa,Osborn:2016bev,Gliozzi:2016ysv,Gliozzi:2017hni}, 
  and the recent  conjectures in \cite{Gadde:2020nwg,Gadde:2023daq}, that conformal field theories with a stress tensor
in  higher dimensions 
are trivial, or non-unitary. 
  The large spin exploration is natural due to the  observation in \cite{Iliesiu:2018fao},
   that thermal one-point functions with large spin are universal. Here we see that these one-point functions asymptote 
   to their Stefan-Boltzmann values at large spin.

To study the dependence of the one-point functions on spin  $l$ and dimension $d$, we chose the ratio
of a given one-point function of spin $l$ operator at the non-trivial fixed point of say the $O(N)$ model or the 
Gross-Neveu model  to Gaussian fixed point in $d = 2k +1$ dimensions.
We denote this ratio by 
\begin{equation}\label{defratio}
r(l, d) = \frac{a_{\cal O}[l]_{\rm m_{\rm th} \neq 0 } }{ a_{\cal O}[l]_{m_{\rm th }=0 }} , 
\qquad l =2, 4, \cdots.
\end{equation}
Note that setting $m_{\rm th} =0$ takes the one-point function to the free theory or the Stefan-Boltzmann result, while we define the
non-trivial fixed point by choosing the value of $m_{\rm th}$ which satisfies the gap equation. 
This ratio  is analogous to the famous ratio  between the stress tensor of ${\cal N}=4$ super-Yang-Mills at strong coupling 
and the Stefan-Boltzmann result which  is $3/4$ or the ratio between the stress tensor of the critical $O(N)$ model at strong coupling 
to the Stefan-Boltzmann result with is $4/5$ for $d=3$. 
Here we examine the ratio at  arbitrary spins not just $l=2$.  

For the critical $O(N)$ model at large $N$, a real solution to the gap equation exists in $2k+1$ dimensions with $k$ odd. 
The ratio (\ref{defratio}) is always less than unity, and 
as $k$ is increased with the spin $l$ held fixed, the ratio  vanishes. 
The same behaviour is seen for the critical Gross-Neveu model at large $N$ which has a real solution to the gap equation 
in $2k+1$ dimensions with $k$ even. 
The fact that on  increasing the dimensions  the ratio vanishes seems to indicate that the number of degrees of freedom 
at the non-trivial fixed point decreases.  It will be interesting to see if such behaviour is true in general not just for the models
studied in this paper. 
When the dimension is fixed and the spin $l$ is increased, we see that for both models the ratio
(\ref{defratio}) tends to unity. 
This is consistent 
with the  perturbative analysis of \cite{Iliesiu:2018fao}, for one-point functions at large spin. 
Their analysis isolated a universal contribution to the one-point functions at large spin.

The organization of the paper is as follows. In the  section \ref{secinv} we discuss the OPE expansion of the two-point function 
of spinor operators and briefly review the OPE inversion formula. 
In section \ref{secmft} we apply our formalism to the MFT of fermions and then in section \ref{secgnmodel} to the 
Gross-Neveu model to obtain thermal one-point functions using the OPE inversion formula. 
In section \ref{seclargds} we study
 the behaviour of the one-point functions of both the critical $O(N)$ model and the Gross-Neveu model  both at large $d$ and at large $l$. 
 Section \ref{secconc}  contains the conclusions . 
 The appendix  \ref{appendgap}
 provides the derivation of the gap equation for the Gross-Neveu model and its stress tensor from the partition function.

\section{Inversion formula for fermionic operators} \label{secinv}

In this section we wish to obtain the Euclidean inversion formulae for CFT's at finite temperature 
with only fermionic operators generalising the 
discussion of \cite{Iliesiu:2018fao}.   
Consider the following fermion bi-linears
\begin{eqnarray} \label{fermbilinear}
{\cal O}_0 &=& \bar \psi \partial^{\mu_1} \cdots \partial^{\mu_J} \partial^{2n}  \psi  - {\rm Traces}, \qquad l = J,   \\ \nonumber
{\cal O}_+ &=& \frac{1}{J+1} \left( 
\bar\psi \gamma_\mu \partial^{\mu_1} \cdots \partial^{\mu_J}   \partial^{2n} \psi  + {\rm cyclic} \right) -{\rm Traces},  \qquad l = J+1,\\ \nonumber
 {\cal O}_- &=& \bar\psi \gamma^\mu \partial_\mu \partial^{\mu_1} \cdots  \partial^{\mu_{J-1}} \partial^{2n} \psi   - {\rm Trace} , \qquad l = J-1,
 \end{eqnarray}
 where $n =0,  1, \ldots$, $l$ is the spin and $J$ the number of derivatives and $\bar \psi = \psi^\dagger$. 
 These are the possible  symmetric traceless tensors formed out of bi-linears of fermions which 
 can have non-trivial expectation value in the thermal vacuum. 
 In this section we obtain 
 the Euclidean inversion formula  which relates  one-point functions of  the above fermion bi-linears 
to the two-point function of the fermions. 
We test the inversion formula by considering the mean field theory of fermions. 
We then apply it 
 the Gross-Neveu model  at large $N$ to derive the one-point functions at finite temperature for the operators listed in 
 (\ref{fermbilinear}).

\subsection{OPE expansion of fermionic correlators} 

Consider the  following  two-point functions  in a CFT at finite temperature
\begin{eqnarray} \label{corr}
g_1 (x) &=& \langle\bar  \psi( x) \psi (0) \rangle_{S^1_\beta \times R^{d-1} } , 
\\ \nonumber
g_2(x) &=& \langle \bar \psi (x) \frac{ \gamma^\mu x_\mu }{|x|} \psi(0) \rangle_{S^1_\beta \times R^{d-1} } , 
\\ \nonumber
g_3(x) &=& \langle \partial_\mu \bar  \psi (x) \gamma^\mu \psi(0) \rangle_{S^1_\beta \times R^{d-1} },
\end{eqnarray}
where $x = ( \tau, x^1 \cdots x^{d-1})$ and $|x|^2 = \ \tau^2 + (x^1)^2 +\cdots (x^{(d-1)})^2 $. 
We will derive inversion formulae relating these two-point functions to the one-point functions in (\ref{fermbilinear}). 
To this we would need the OPE expansions of these correlators. 
Let the OPE of the fermion bi-linear be given by 
\begin{equation} \label{ope}
\psi^{\dagger}_\alpha (x) \psi_\beta (0) = \sum_{{\cal O} \in \psi^\dagger \times \psi } 
\frac{ f_{\psi^\dagger \psi {\cal O}} }{c_{\cal O}} |x|^{\Delta_{\cal O}  -2\Delta_{\psi} -J} x_{\mu_1} \cdots x_{\mu_J} 
{\cal O}^{ \mu_1\cdots \mu_J}_{\beta\alpha}  (0)  + \cdots.
\end{equation}
Here ${\cal O}^{\beta \alpha \mu_1\cdots \mu_J} (0) $ are all the operators that occur in the OPE of the 
fermions. The tensor indices are symmetric and traceless. 
The $\cdots$ refers to other representations which are anti-symmetric in any pair of the 
tensor indices. These vanish   in the thermal vacuum and therefore not relevant for our purpose. 
The representations  of $SO(d)$  which acquire non-trivial expectation values in the vacuum $S^1_\beta \times R^{d-1}$ 
are those which contain the trivial representation under $O(d-1)$.
The fermion bi-linear indices in (\ref{ope})  together with the 
tensor indices can be combined into irreducible representations of $SO(d)$, we will do this subsequently for each 
of the correlators.
The coefficients $f_{\psi^\dagger \psi {\cal O}}$, are the structure constants and $c_{\cal O}$ is the 
normalization of the two-point function of the operator ${\cal O}$.

Let us
substitute the OPE  (\ref{ope}) into the correlators given in (\ref{corr}).
For the first correlator we obtain 
\begin{equation} \label{g1}
g_1 (x) = \sum_{{\cal O} \in \psi^\dagger \times \psi } 
\frac{ f_{\psi^\dagger \psi {\cal O}} }{c_{\cal O}} |x|^{\Delta_{\cal O}  -2\Delta_{\psi} -J} x_{\mu_1} \cdots x_{\mu_J}
\langle {\cal O}^{ \alpha; \;\; \mu_1\cdots \mu_J}_{\;\alpha} \rangle.
\end{equation}
The thermal expectation values on  the  right hand side  of (\ref{g1}) 
are the one-point functions which we are interested in computing. 
In a theory of only fermions, 
this class of operators can be written as the fermion bi-linears \footnote{In general these operators are just  bosonic
traceless symmetric tensors.   For instance if there are Yukawa couplings in the theory
  they could be also be  made of bosonic bi-linears. In this work we will restrict our attention to theories without such couplings.}
\begin{equation}
{\cal O}_0^{\mu_1\cdots \mu_J} \equiv {\cal O}^{\alpha;  \mu_1\cdots \mu_J}_{\;\alpha}  = \psi^\dagger \partial^{\mu_1}\cdots \partial^{\mu_J}  \psi - {\rm traces} .
\end{equation}
Using  translational invariance and spatial rotational invariance 
 of the thermal vacuum, we have the following result for thermal one-point functions of  symmetric traceless tensors.
\begin{equation} \label{tproperty}
\langle {\cal O}^{\mu_1\cdots \mu_J} (x) \rangle =  b_{\cal O} T^{\Delta_{\cal O} } 
( e^{\mu_1} e^{\mu_2} \cdots e^{\mu_J}  - {\rm Traces} ) .
\end{equation}
Here $e^\mu$ is the unit vector in the thermal direction $\tau$. 
Now we also have the identity
\begin{equation} \label{iden}
|x|^{-J} (x_{\mu_1} \cdots x_{\mu_J} ) ( e^{\mu_1} e^{\mu_2} \cdots e^{\mu_J}  - {\rm Traces} )
=\frac{J!}{2^J (\nu)_J} C_J^{(\nu) } (\eta) ,
\end{equation}
where
\begin{equation}
\nu = \frac{d-2}{2}, \qquad ( a)_n = \frac{\Gamma( a+ n) }{\Gamma (a) }, 
\qquad \eta = \frac{\tau}{|x|},
\end{equation}
and $C_J^{(\nu) } (\eta) $ is the Gegenbauer polynomial of degree $J$. 
Using the property (\ref{tproperty}) and (\ref{iden}) in the expression (\ref{g1}), we obtain 
\begin{eqnarray} \label{opexp1}
g_1 (x) &=& \sum_{{\cal O} \in \psi^\dagger \times \psi }  |x|^{\Delta_{{\cal O} } - 2\Delta_{\psi}}  a_{{\cal O}_0} C_{J}^{(\nu)} (\eta), 
\\ \nonumber
a_{{\cal O}_0}&=&  b_{{\cal O}_0} T^{\Delta_{\cal O}} \frac{J!}{2^J (\nu)_J} \frac{ f_{\psi^\dagger \psi {\cal O} }}{c_{\cal O} }.
\end{eqnarray}
The above equation is the OPE expansion of the  finite temperature  two-point function $g_1(x)$,   in terms of 
thermal one-point functions of fermion bi-linears $a_{{\cal O}_0}$.

Let us now repeat the analysis for  the two-point function $g_2(x)$.  Substituting the OPE (\ref{ope}), we obtain 
\begin{equation} \label{g2}
g_2(x) =  \sum_{{\cal O} \in \psi^\dagger \times \psi }   |x|^{\Delta_{\cal O} -2\Delta_{\psi} -J-1} x_{\mu_1} \cdots x_{\mu_J} x_\nu
\langle\gamma^{\nu \; \alpha\beta} {\cal O}^{ \mu_1\cdots \mu_J}_{\beta \alpha} \rangle.
\end{equation}
Now the operator on the RHS can be decomposed into various  irreducible representations,
\begin{eqnarray} \label{irrep}
\gamma^{\nu \; \alpha\beta} {\cal O}^{ \mu_1\cdots \mu_J}_{ \beta\alpha} &=&
\left[ \frac{1}{J +1} \left(  \gamma^{\nu \; \alpha\beta} {\cal O}^{ \mu_1\cdots \mu_J}_{ \beta\alpha}  + 
 \gamma^{\nu_1 \; \alpha\beta} {\cal O}^{ \mu_2\cdots \mu_J\mu}_{ \beta\alpha} + {\rm cyclic}  \right) 
 -{\rm Traces}  \right] + {\rm Traces} \nonumber \\ 
 && +  \frac{1}{J +1}  \left(  \gamma^{\nu \; \alpha\beta} {\cal O}^{ \mu_1\cdots \mu_J}_{ \beta\alpha} - 
 \gamma^{\nu_1 \; \alpha\beta} {\cal O}^{ \mu \mu_2 \cdots \mu_J}_{ \beta\alpha}  \right) 
 + \cdots ( J-1 ){\rm terms} )  ,
\end{eqnarray}
where `Traces'   are the terms subtracted to ensure that the term in the square bracket on the first line 
is a rank $J+1$ traceless symmetric tensor. 
The `Traces' are given by 
\begin{equation}
{\rm Traces} = \frac{2}{(J+1)(d + J-1)} 
\left( \delta^{\mu\mu_1} \gamma^{\rho\;\alpha\beta} {\cal O}^{ \rho_{\mu_2} \cdots \mu_J}_{ \beta\alpha} 
+ \delta^{\mu_1\mu_2} \gamma^{\rho\;\alpha\beta} {\cal O}^{ \rho_{\mu_3} \cdots \mu_J \mu_1 }_{ \beta\alpha} 
+ {\rm cyclic} \right) .
\end{equation}
The equation (\ref{irrep}) essentially writes the tensor product of a vector with a symmetric  traceless tensor of 
rank $J$  as a sum  of symmetric traceless  tensors of rank $J+1$ and rank $J-1$ together with tensors which are 
anti-symmetric in two of the indices. 
Now since thermal expectation values are non-zero only for symmetric tensors,  the tensors which are anti-symmetric 
in any two of the indices can be ignored.
Using the property (\ref{tproperty}) and the identity (\ref{iden}) in the  expression the correlator $g_2(x)$ given in 
(\ref{g2}) we obtain 
\begin{equation} \label{g2expope}
g_2(x) =  \sum_{{\cal O} \in \psi^\dagger \times \psi } |x|^{\Delta_{\cal O} - 2\Delta_{\psi} } 
\left(  a_{{\cal O}_+} C^{(\nu)}_{J+1} (\eta)  + \frac{ 2J}{(J+1)( d+ J-1)} a_{{\cal O}_{-} } C^{(\nu)}_{J-1}(\eta)  \right),
\end{equation}
where 
\begin{eqnarray}
a_{{\cal O}_{+}} = b_{{\cal O}_+} T^{\Delta_{{\cal O}_+}}
 \frac{(J+1)! }{2^{J+1} (\nu)_{J+1} } \frac{ f_{\psi^\dagger \psi {\cal O} }}{c_{\cal O}}, 
\\ \nonumber
 a_{{\cal O}_{-}} = b_{ {\cal O}_- } T^{\Delta_{{\cal O}_-} }
 \frac{(J-1)!}{2^{J-1} (\nu)_{J-1} } \frac{ f_{\psi^\dagger \psi {\cal O} } }{c_{\cal O} }.
\end{eqnarray}
Here again we restrict our attention to the case where the operators on the R.H.S of (\ref{g2expope}) are 
 fermion bi-linears given by 
\begin{eqnarray}
{\cal O}_+ &=&\frac{1}{J+1}\left( 
 \psi^\dagger \gamma^\mu \partial^{\mu_1} \cdots \partial^{\mu_J} \psi + {\rm cyclic} \right)  - ( {\rm Traces}), 
 \\ \nonumber
 {\cal O}_- &=& \psi^{\dagger }\gamma^{\mu } \partial_\mu \partial^{\mu_1} \cdots \partial^{\mu_{J-1}}  \psi.
\end{eqnarray}

Finally let us examine the correlator $g_3(x)$. From (\ref{ope}), we obtain  the OPE
\begin{eqnarray}
\partial_\mu \psi^\dagger(x) \gamma^\mu \psi &=& \sum_{{\cal O} \in \psi^\dagger \times \psi }
\frac{f_{\psi^\dagger\psi {\cal O}} }{c_{\cal O}}  \left( 
|x|^{ \Delta_{\cal O} - 2\Delta_{\psi} - J -2} ( \Delta_{\cal O} - 2\Delta_{\psi} - J) x_\mu x_{\mu_1} \cdots x_{\mu_{J}}
{\cal O}_+^{\mu \mu_1 \cdots x_{\mu_J}}  \right. \nonumber \\ \nonumber
&& \left. + 
\Big( J + \frac{2J( \Delta_{\cal O} - 2\Delta_{\psi} - J)}{(J+1)( J+ d-1) } \Big )  |x|^{ \Delta_{\cal O} - 2\Delta_{\psi} - J}  x_{\mu_1} \cdots x_{\mu_{J-1}}
{\cal O}_-^{ \mu_1 \cdots x_{\mu_{J-1}} } + \cdots  \right).  \\
\end{eqnarray}
We can now take thermal expectation values and obtain 
\begin{eqnarray} \label{opexp3}
g_3(x) &=& \sum_{{\cal O} \in \psi^\dagger \times \psi }|x|^{ \Delta_{\cal O} - 2\Delta_{\psi} -1}
\left[ ( \Delta_{\cal O} - 2\Delta_{\psi} - J) a_{{\cal O}_+} C^{(\nu)}_{J+1} (\eta)   \right. \\ \nonumber
&& \qquad\qquad\qquad \left.  +\Big( J + \frac{2J ( \Delta_{\cal O} - 2\Delta_{\psi} - J)}{(J+1)( J+ d-1) } \Big ) a_{{\cal O}_-}  C^{(\nu)}_{J-1} (\eta) 
\right].
\end{eqnarray}

It is important to realise that due to the presence of operators belonging to the class ${\cal O}_+$ as well as 
${\cal O}_-$ in the OPE expansions of $g_2(x)$ and $g_3(x)$ given in (\ref{g2expope}) and  (\ref{opexp3}), the OPE inversion 
formulas for the one-point functions will involve both these correlators.

\subsection{Euclidean inversion formulas} \label{eucinv}

In this section we  briefly review the Euclidean inversion formula introduced in 
\cite{Iliesiu:2018fao} and  obtain the expressions relating the 
one-point functions listed in (\ref{fermbilinear}) to the thermal 2-point functions. 
One difference we need to keep track is the fact that the $2$ point functions 
$g_2(x)$ and $g_3(x)$ involve a linear combination of 
one-point functions of operators belonging to class ${\cal O}_+$ and ${\cal O}_-$ and therefore 
the inversion formula for these operators will involve linear combinations of $g_2(x)$ and $g_3(x)$.

Consider the OPE expansion of a correlator  given in the form
\begin{equation} \label{genope}
g(x) = \sum_{{\cal O}} |x|^{\Delta_{\cal O}  - 2\Delta_{\psi} } a_{\cal O} C_{l}^{(\nu)} (\eta) .
\end{equation}
The expansions in (\ref{opexp1}),  (\ref{g2expope}) and (\ref{opexp3}) are of this form. 
By introducing  the spectral function $\hat a(\Delta, l)$ we can write the OPE expansion 
in terms of an integral
\begin{equation} \label{contrep}
g(x) = \sum_{l=0}^\infty \oint_{-\epsilon -i \infty}^{-\epsilon + i \infty} \frac{d\Delta}{ 2\pi i } 
\hat a(\Delta, l) C_{l}^{(\nu)} ( \eta) |x|^{\Delta_{} - 2 \Delta_\psi} .
\end{equation}
Here the spectral function should have poles of the form 
\begin{equation}\label{locpoles}
\hat a(\Delta, l) \sim -\frac{a_{\cal O}}{ \Delta -\Delta_{\cal O}}.
\end{equation}
The contour in (\ref{contrep}) is chosen to encircle the right-half of the $\Delta$ plane when $|x|<1$ and
demanding the spectral function does not grow exponentially in this region.
Deforming the contour to encircle the poles results in the sum given in (\ref{genope}). 
The contour in (\ref{contrep}) has been chosen so that all the physical poles along with the unit operator are included. 

We can invert the equation (\ref{contrep}) using the orthogonality of the Gegenbauer polynomials.
\begin{equation} \label{laplace} 
\hat a(\Delta_{}, l)  =\frac{1}{N_l} \int_{|x|<1}  d^dx C_{l}^{(\nu)}(\eta)  |x|^{2\Delta_\psi -\Delta_{} - d} g(x).
\end{equation}
Here we first use the property
\begin{eqnarray}
\int_{S^{d-1}}  d\Omega C_l^{(\nu)}(\eta) C_{l'}^{(\nu)} (\eta) =N_l \delta_{ll'}, \\ \nonumber
N_l = \frac{4^{1-\nu} \pi^{\nu + \frac{3}{2} } \Gamma( l+ 2\nu) }{ l! (l+\nu) \Gamma(\nu)^2 \Gamma(\nu + \frac{1}{2}) },
\end{eqnarray}
to fix on to a particular $l$.
Then the integral over $x$ functions as the Laplace transform which picks out the relevant pole.
 It can be seen that  (\ref{laplace}) is consistent,
by substituting for $g(x)$ from (\ref{contrep}). 

Now the Euclidean inversion formula (\ref{laplace}) is cast as an integral over the 2 dimensional plane using 
rotational invariance.
Let us first discuss the case $d>2$.  Using  the spatial $SO(d-1)$ rotational invariance we can choose to write 
the $d$ component vector $x$ as $x = ( \tau, x_E, 0, \cdots )$.  So the relevant kinematics can be parametrized 
by introducing the following  complex variables, as well as polar coordinates.
\begin{eqnarray} \label{parem}
z= \tau + i x_E, \qquad \bar z = \tau - i x_E, \\ \nonumber
z = rw, \qquad \bar z = rw^{-1}.
\end{eqnarray}
Note that in these variables
\begin{equation}
\eta = \frac{\tau}{|x| } = \cos \theta = \frac{1}{2} ( w + w^{-1}).
\end{equation}
Therefore the  Gegenbauer polynomials are   functions of the polar angle $\theta$. 
It can be written in terms of the hypergeometric function 
\begin{equation}\label{genhyp}
C_l^{(\nu)} \Big( \frac{1}{2} ( w + w^{-1}) \Big) = 
\frac{\Gamma( l+ 2\nu)} { \Gamma( \nu ) \Gamma( l + \nu +1) }
 \Big( F_l(w^{-1} ) e^{i\pi \nu}+ F_l( w) e^{ - i \pi \nu } 
 \Big), \qquad {\rm Im}\, w>0
\end{equation}
where 
\begin{equation} \label{genhyp1}
F_l(w) = w^{l+ 2\nu } {}_2 F_1( l + 2\nu , \nu , l + \nu +1, w^2) .
\end{equation}
The representation in terms of hypergeometric function allows to continue $w$ to the entire complex plane. 
For ${\rm Im}\, w<0$, the phases of the two terms in (\ref{genhyp}) are exchanged. 
We see using (\ref{parem}), that  $g(x)$ is a function of $g(z, \bar z)= g (r, \theta) $.  
These observations 
allow us to perform all the remaining $d-2$ angular integrals in (\ref{laplace}) leaving the integral over
the complex plane $(z, \bar z)$.  

Consider $g(z, \bar z) = g(rw, rw^{-1})$ as a function in the complex $w$ plane. 
We assume the following analytic properties in the $w$ plane \cite{Iliesiu:2018fao}:
 The 2 point function is analytic in the $w$ plane except at the branch cuts $(-\infty , -1/r)$, $ (-r, 0)$, $(r, 0)$, $(1/r, \infty)$. 
 The second assumption is 
that at large $w$ the growth of $g(rw, rw^{-1})$ is bounded  by the polynomial $w^{l_0}$ for a fixed $l_0$. 
At small $w$ the growth of  $g(rw, rw^{-1})$  is bounded by  $w^{-l_0}$. 
These 2 properties allow one to deform the integral contour over $w$ along the branches
$(-\infty,-\frac{1}{r})\cup(\frac{1}{r},\infty)$, together with the circle at $\infty$  from that of the unit circle.\footnote{The integration contour can also be deformed towards the origin and this again can be related to the integral along the contour deformation mentioned above as it is illustrated in detail in \cite{Iliesiu:2018fao}.}
Using these methods and a change of variables the inversion formula can be written as
\begin{eqnarray}\label{finalinversion}
\hat a( \Delta_{} , l) &=&  \hat a_{\rm disc} ( \Delta_{}, l) + \theta( l_0 -l) \hat a_{\rm arcs} ( \Delta_{} ,l), \\ \nonumber
\hat a_{\rm disc} ( \Delta_{}, l)&=& ( 1 + (-1)^l) K_l
 \int_0^1\frac{d\bar z}{\bar z} \int_1^{\frac{1}{\bar z}} \frac{dz}{z} (z\bar z)^{\Delta_{\psi} -\frac{\Delta_{}}{2} -\nu}
 (z-\bar z)^{2\nu}
F_l \Big( \sqrt{\frac{\bar z}{z}} \Big) {\rm Disc} [g(z, \bar z) ], \\ \nonumber
K_l &=& \frac{\Gamma(l+1) \Gamma(\nu) }{ 4\pi \Gamma(l+\nu) }.
\end{eqnarray}
Here  the discontinuity across the branch cuts is given by 
\begin{equation}
{\rm Disc}[g(z, \bar z) ] = \frac{1}{i} \big( g( z +i \epsilon , \bar z) - g( z-i\epsilon, \bar z) \big) .
\end{equation}
For $l<l_0$ the contribution from the arcs, which essentially becomes an integral over the circle at infinity 
is given by the term $ \hat a_{\rm arcs} ( \Delta_{} ,  l)$ which needs to be evaluated by performing the following   integral in
(\ref{finalinversion})  over the circle
at infinity in the $w$-plane explicitly. 
\begin{eqnarray}\label{db2arc}
\hat a_{\rm arcs} (\Delta_{},  l)  &=&  2 K_l   \int_0^1 \frac{dr}{r^{\Delta_{}  +1 - 2\Delta_\psi} }  \times  \\ \nonumber
&& \oint \frac{dw}{i w} \lim_{|w| \rightarrow \infty} 
\left[ \Big( \frac{ w - w^{-1} }{i} \Big)^{2\nu} 
 F_l(w^{-1}) e^{i\pi\nu}  g( r, w) 
\right].
\end{eqnarray}
The expression in (\ref{finalinversion}) is the  form of the inversion formula we will use for $d>3$. Note that 
it can be applied to the three correlators in (\ref{corr}) as their OPE expansions 
(\ref{opexp1}), (\ref{g2expope}) and (\ref{opexp3}) are of the 
form (\ref{genope}). 

For $d=2$, we need to treat the normalization of the Gegenbauer polynomials carefully. 
but in the end the 
inversion formula is very similar. 
It is best to first re-examine the  OPE representation of the thermal two-point function 
which can be written explicitly as 
\begin{equation} \label{opexp2d}
g(x) = \sum_{\cal O} |x|^{\Delta_{\cal O} - 2 \Delta_{\psi} } b_{\cal O} T^{\Delta_{\cal O}} \frac{l!}{2^l ( \nu)_l} \frac{f_{\psi^\dagger \psi {\cal O}}}{c_{\cal O}} C_l^{(\nu)} ( \eta) .
\end{equation}
In the limit $d\rightarrow 2$ or $\nu\rightarrow 0$, 
the Gegenbauer polynomials take the limiting form
\begin{eqnarray} \label{limgegenb}
\lim_{\nu\rightarrow 0} C_l^{(\nu)} (\eta) &=& \frac{2\nu}{l} \Big( \frac{ w^l + w^{-l}}{2} \Big) =
  \frac{2\nu}{l} \cos ( l \theta) ,  \qquad l >0, \\ \nonumber
  \lim_{\nu \rightarrow 0} C_0^{(\nu)} (\eta) &=& 1.
\end{eqnarray}
From (\ref{genhyp1}) we also obtain the following limit
\begin{equation}
\lim _{\nu\rightarrow 0} F_l(w)  = w^l.
\end{equation}
Taking the limit $\nu\rightarrow 0$ in  (\ref{opexp2d}), we obtain
\begin{eqnarray} \label{2drelation}
&&g(x) = \sum_{\cal O} |x|^{\Delta_{\cal O} - 2 \Delta_{\psi} } {a_{\cal O}}|_{d=2}, \qquad
a_{\cal O}|_{d=2} =  \frac{ b_{\cal O} }{2^{l-1} } T^{\Delta_{\cal O} }  \frac{f_{\psi^\dagger \psi {\cal O} }}{c_{\cal O}} , 
\\ \nonumber
&&  \qquad a_{\cal O}|_{d=2} = \lim_{\nu\rightarrow 0} \frac{2\nu }{l}  a_{\cal O}|_{\nu}, \; {\rm for} \; l>0, 
\qquad a_{\cal O}|_{d=2}  =  \lim_{\nu\rightarrow 0 }  a_{\cal O}|_{\nu}, \;\; {\rm for}\; l =0.
\end{eqnarray}
Going through a similar analysis we obtain 
\begin{equation}
\hat a(\Delta_{}, l)_{\rm disc}  |_{d=2}  = ( 1 + (-1)^l) \frac{1}{2\pi} \int_0^1 \frac{d\bar z}{\bar z} \int_1^{\frac{1}{\bar z}}
\frac{dz}{z} z^{\Delta_{\psi} - \bar h} \bar z^{\Delta_{\psi} -h }
{\rm Disc}[ g(z, \bar z)],
\end{equation}
with
\begin{equation}
h = \frac{\Delta_{}  -l}{2}, \qquad {\rm and}\quad  \bar h =  \frac{\Delta_{}  +l}{2}.
\end{equation}
The contribution from the arcs is given by 
\begin{equation}
\hat a_{\rm arcs} ( \Delta_{}, l)|_{d=2} = \frac{1}{2 \pi} \int_0^1
 \frac{1}{ r^{\Delta_{} +1 - 2\Delta_\psi}} \oint \frac{dw}{i w} \lim_{|w|\rightarrow\infty} w^{-l} .
\end{equation}

\section{Mean field theory of fermions} \label{secmft}

In mean field theory, a $2n$ point function is given by pairwise contraction of the $n$ two-point functions. 
Therefore, the  $2$ point function at finite temperature can be obtained by using the method of images. 
Consider a fermionic operator $\psi$ of dimension $\Delta_\psi$ in MFT, then the thermal two-point function 
is given by 
\begin{equation} \label{gnftcor}
\langle \psi_{\alpha} (x) \psi^\dagger_\beta(0) \rangle_\beta 
= \frac{1}{2^{[\frac{d}{2}]}}
\sum_{m\in\mathbb{Z}} (-1)^m \frac{\gamma_{\alpha\beta}^\mu x_\mu^{(m)}}{|x^{(m)}|^{2\Delta_\psi+1}}, \qquad\text{where} \qquad x^{(m)}_\mu\equiv\{(\tau+m),\vec{x}\} .
\end{equation}
We have set the inverse temperature $\beta =1$. This correlator obeys anti-periodic boundary conditions 
along the thermal circle.  We have normalized the correlator, by  $2^{[\frac{d}{2}]}$, the dimension of the 
Dirac spinor.  
This is for convenience so that this factor cancels on taking the trace over the $\gamma$-matrices. 
We will account for this while comparing with our results from the partition function. 
Note that scalar correlator $g_1(x)$ vanishes, this implies in MFT the class of operators ${\cal O}_0$ has zero expectation 
value in the thermal vacuum. 

\subsubsection*{$g_2(\tau, \vec x)$}
Let us evaluate the correlator $g_2(x)$ for MFT \footnote{ We have kept track of the ordering of the $\psi$ and $\bar \psi$, which is different in the definition of $g_2(x)$ in  (\ref{corr}).}
\begin{align} \label{g2mft}
	g_2(\tau, {\vec x} ) =\frac{1}{|x|^{2 \Delta_\psi}}+\sum_{\substack{m=-\infty\\m\ne0}}^{\infty}(-1)^m\frac{m\tau+|x|^2}{[(\tau+m)^2+\vec{x}^2]^{\Delta_\psi+\frac{1}{2}} \,|x|}.
\end{align}
We can  systematically expand this correlator in small $x$  so as to 
compare with the 
OPE expansion in (\ref{g2expope}).  This expansion is facilitated by the identity
\begin{equation}
\frac{1}{ (1- 2xy + y^2)^\alpha} = \sum_{j=0}^\infty C_j^{(\alpha)} y^j,
\end{equation}
where $C_j^{(\alpha)}$ are Gegenbauer polynomials of order $j$ with index $\alpha$. 
Once the OPE expansion and the thermal expectation value of the operators in the class ${\cal O}_+$, ${\cal O}_-$ are  obtained, 
they can be compared against the same obtained from the inversion formula. 
This will provide an important check on the inversion formula.  Indeed, when the formula is applied to $g_2(\tau, x)$ given in 
(\ref{g2mft}), this case would be a distinct check from that done in \cite{Iliesiu:2018fao}. 
Proceeding with the expansion we obtain
\begin{align} \label{mang2}
g_2(\tau, \vec{x}) &=\frac{1}{|x|^{2 \Delta_\psi}}+\sum_{\substack{m=-\infty\\m\ne0}}^{\infty}(-1)^m\left(m\eta+|x|\right) \times \sum_{j=0}^\infty (-1)^j C_j^{(\Delta_\psi+\frac{1}{2})}(\eta)\, \frac{{\rm sgn}(m)^j |x|^j}{|m|^{2\Delta_\psi+1+j}},\nonumber\\
	&=\frac{1}{|x|^{2 \Delta_\psi}}-\sum_{j=1,3,\ldots}2\eta\,C_j^{(\Delta_\psi+\frac{1}{2})}(\eta)|x|^j (2^{1-2\Delta\psi-j}-1)\zeta{(2\Delta_\psi+j)} \nonumber\\ 
	&\qquad+\sum_{j=0,2,4,\ldots}2\,C_j^{(\Delta_\psi+\frac{1}{2})}(\eta)|x|^{j+1}(2^{-2\Delta\psi-j}-1)\zeta({2\Delta_\psi+1+j}).
\end{align}
In the second line we have cancelled off terms which occur with equal and opposite signs and then performed the 
sum over $m$.  The recurrence  relation 
\begin{equation}
2 (n+ \lambda) \eta C^{(\lambda)} (\eta) = 
(n+1) C_{n+1}^{(\lambda)} (\eta) + ( n-1 +2 \eta) C^{(\lambda)}_{n-1}( \eta) ,
\end{equation}
can be used to remove the explicit factor of $\eta$ in the second line of (\ref{mang2}). 
This results in 
\begin{align}
	&g_2(\tau, \vec x)=\frac{1}{|x|^{2 \Delta_\psi}}+\sum_{j=0,2,4,\ldots}2\, |x|^{j+1}(2^{-2\Delta_\psi-j}-1)\zeta({2\Delta_\psi+1+j}) C_j^{(\Delta_\psi+\frac{1}{2})} (\eta)\nonumber\\ 
	&-\sum_{j=1,3,\ldots}2|x|^j \frac{(2^{1-2\Delta\psi-j}-1)}{2(j+\Delta_\psi+\frac{1}{2})}\zeta{(2\Delta_\psi+j)} \bigg((j+1)C_{j+1}^{(\Delta_\psi+\frac{1}{2})}(\eta)+(j+2\Delta_\psi)C_{j-1}^{(\Delta_\psi+\frac{1}{2})}(\eta)\bigg).
\end{align}
To compare with the OPE expansion of the correlator, we would need the index of the Gegenbauer polynomials 
to be $\nu$ instead of $\Delta_{\psi} + \frac{1}{2}$. For this we can use the following identity
\begin{align}
	C_j^{(\Delta)}(\eta)=\sum_{l=j,j-2,\ldots,j\, \text{mod }2}\frac{(l+\nu)(\Delta)_{\frac{j+l}{2}}{(\Delta-\nu)_{\frac{j-l}{2}}}}{(\frac{j-l}{2})!(\nu)_{\frac{j+l+2}{2}}} C_l^{(\nu)} (\eta).
\end{align}
Then expressing $j = 2n +l$ and grouping terms with same summation ranges we obtain
\begin{align}
	&	g_2(\tau, \vec x) =\frac{1}{|x|^{2 \Delta_\psi}}+\nonumber\\
	&	\sum_{\substack{n=0\\l=0,2,\ldots}}^\infty2|x|^{2n+l+1}(2^{-2\Delta_\psi-2n-l}-1)\zeta({2\Delta_\psi+1+2n+l})\bigg(1-\frac{2n+l+1+2\Delta_\psi}{2(2n+l+\Delta_\psi+\frac{3}{2})}\bigg)\nonumber\\
	&\hspace{6cm}
	\times\frac{(l+\nu)(\Delta_\psi+\frac{1}{2}-\nu)_n(\Delta_\psi+\frac{1}{2})_{n+l}}{n!(\nu)_{n+l+1}} C_l^{(\nu)}(\eta)\nonumber\\
	&
	-\sum_{\substack{n=0\\l=0,2,\ldots\\n+l\ne0}}^\infty2|x|^{2n+l-1} \frac{(2^{2-2\Delta_\psi-2n-l}-1)}{2(2n+l+\Delta_\psi-\frac{1}{2})}\zeta{(2\Delta_\psi+2n+l-1)} (2n+l)\nonumber\\
	&\hspace{6cm}\times\frac{(l+\nu)(\Delta_\psi+\frac{1}{2})_{n+l}(\Delta_\psi+\frac{1}{2}-\nu)_{n}}{n!(\nu)_{n+l+1}} C_l^{(\nu)}(\eta).
\end{align}
To compare with the OPE expansion, it is useful to separate the $n=0, l =2, 4, \cdots$ terms from the last line of the above equation. 
Then combine the rest with the terms in the first summation. 
We can group these terms by shifting $n\rightarrow n-1$ in the first summation. 
These manipulations lead to 
\begin{align}\label{g2expans}
	&	g_2(\tau, \vec x) =\frac{1}{|x|^{2 \Delta_\psi}}+\nonumber\\
	&\sum_{\substack{n=1\\l=0,2,\ldots}}^\infty\frac{2^{}(l+\nu ) (l+2 n)  \left(1-2^{-2 \Delta_\psi-l-2n+2}\right) 
	 \zeta (l+2 n+2 \Delta_\psi -1)  | x| ^{l+2 n-1}C_l^{(\nu)}(\eta)}{ \Gamma (n+1) (2 \Delta +2 l+4 n-1)  (\nu )_{l+n+1}}\nonumber\\
	&\qquad\times  \big(\Delta_\psi +\frac{1}{2}\big)_{l+n}  \big(\Delta_\psi -\nu +\frac{1}{2}\big)_{n} \bigg(1-\frac{4 n (l+\nu +n)}{(2 \Delta_\psi +2 l+2 n-1) (2 \Delta_\psi -2 \nu +2 n-1)}\bigg)\nonumber\\
	&+\sum_{l=2,4,\ldots}\frac{2l \left(2^{-2 \Delta_\psi -l+2}-1\right) (l+\nu )  \left(\Delta_\psi +\frac{1}{2}\right)_l \zeta (l+2 \Delta_\psi -1) | x| ^{l-1}C_l^{(\nu)}(\eta)}{(-2 \Delta_\psi -2 l+1) (\nu )_{l+1}}.
\end{align}

We re-write the OPE expansion in (\ref{g2expope}) as 
\begin{align}\label{g2ope}
	g_2(\tau, \vec x )= \sum_{n=0}^\infty\frac{|x|^{2n+1}}{d} a_{\mathcal{O}_-}[n,l=0] C_0^\nu(\eta)+\sum_{l=1}^\infty|x|^{l-1}a_{\mathcal{O}_+}[n=0,l]C_l^\nu(\eta)\nonumber\\+\sum_{\substack{n=1\\l=1}}^\infty|x|^{2 n+l-1}\bigg(a_{\mathcal{O}_+}[n,l]+\frac{2(l+1)}{(l+2)(d+l)}a_{\mathcal{O}_-}[n-1,l]\bigg) C_l^\nu(\eta).
\end{align}
Here we have re-labelled the sum over $J$ as $l$ and separated out the $l=0$ term as well as the $n=0$ term. 
We have also used 
\begin{equation} \label{dimoperator}
\Delta_{{\cal O}_+[n, l] }  = 2\Delta_\psi  +l -1 + 2n , \qquad  \Delta_{{\cal O}_-[n, l] }  = 2\Delta_\psi  +l  +1 + 2n ,
\end{equation}
where $l$ refers to the spin of the operators, $J$ refers to the number of derivatives in the operators. 
Now comparing (\ref{g2expans}) and (\ref{g2ope}) we see that only 
 operators with even spins have non-trivial expectation values in MFT. 
We also obtain
\begin{align}\label{ominusl0}
	a_{\mathcal{O}_-}[n,l=0]=\frac{2\nu(\nu+1)  ( 2 \nu +1-2 \Delta_\psi) \left(2^{-2 (\Delta_\psi +n)}-1\right) 
	\left(\Delta_\psi +\frac{1}{2}\right)_n \left(\Delta_\psi -\nu +\frac{1}{2}\right)_n}{n! (\nu +n+1) (\nu )_{n+1}} \nonumber\\
	\times \zeta (2 n+2 \Delta_\psi +1),
\end{align}
and 
\begin{align} \label{oplusn0}
	a_{\mathcal{O}_+}[n=0,l]=\frac{2l \left(2^{-2 \Delta_\psi -l+2}-1\right) (l+\nu )  \left(\Delta_\psi +\frac{1}{2}\right)_l \zeta (l+2 \Delta_\psi -1)}{(-2 \Delta_\psi -2 l+1) (\nu )_{l+1}}, \qquad  l = 2, 4,  6 \cdots .
\end{align}
Finally  for operators with $n\geq 1$ and $ l = 2, 4, 6, \cdots $ we get a single  linear equation
 \begin{align}\label{1st eq}
	&a_{\mathcal{O}_+}[n,l]+\frac{2(l+1)}{(l+2)(d+l)}a_{\mathcal{O}_-}[n-1,l]\nonumber\\
	&=\frac{(l+\nu ) (l+2 n) \Gamma (\nu ) 2^{-l-2 (\Delta_\psi +n)} \left(2^{l+2 (\Delta_\psi +n)}-4\right) \Gamma \left(l+n+\Delta_\psi -\frac{1}{2}\right)  \Gamma \left(n+\Delta_\psi -\nu -\frac{1}{2}\right)}{\Gamma \left(\Delta_\psi +\frac{1}{2}\right) \Gamma (n+1) \Gamma \left(\Delta_\psi -\nu -\frac{1}{2}\right) \Gamma (l+n+\nu +1)}\nonumber\\
	&\hspace{7cm}\times\zeta (l+2 n+2 \Delta_\psi -1).
\end{align}
At this point there are some observations we can make:
Note that  for $\Delta_{\psi} = \nu  + \frac{1}{2}$, the MFT reduces to the theory of  free fermions in $d$ dimensions. 
Therefore by equations of motion we must have 
\begin{eqnarray} \label{freevals}
a_{{\cal O}_-[n, l]}&=&0,  \quad \mbox{for}\; \; {\Delta_{\psi} = \nu  + \frac{1}{2}} \;\; \mbox{and}\;\;  {n =0, 1, \cdots}, \\ \nonumber
a_{{\cal O}_+[n, l]} &=& 0 ,  \quad \mbox{for}\; \; {\Delta_{\psi} = \nu  + \frac{1}{2}} \;\; \mbox{and}\;\;  {n = 1,2,  \cdots}.
\end{eqnarray}
It can be easily seen that (\ref{ominusl0}) satisfies this requirement and (\ref{1st eq}) is consistent with  (\ref{freevals}). 
Next, notice that the class of operators $a_{{\cal O}_+[n =0, l]}$  has the stress tensor, let us examine the one-point
function for the free field case
\begin{equation}
a_{{\cal O}_+[n=0, l]}= -\frac{2l ( 2^{3- (d+l)} -1) }{ (d-2)} \zeta(l-2+ d) \qquad 
\mbox{for}\;\;  {\Delta_{\psi} = \nu  + \frac{1}{2}}  .
\end{equation}
We see that the result is  proportional  to $\zeta(d)$ for $l=2$, which  is the result expected for 
the stress tensor of free fermions.
The reason the above expression has a divergence at $d=2$ is due to the behaviour of the Gegenbauer polynomials 
in the limit $\nu \rightarrow 0$ as shown in (\ref{limgegenb}). 
Using the relation given in (\ref{2drelation})  which relates the one-point function in $d=2$ to that in arbitrary $d$, we obtain 
\begin{equation}
{a_{{\cal O}_+[n=0, l]}}|_{d=2} = -2 ( 2^{1-l} -1)  \zeta(l) \qquad 
\mbox{for}\;\;  {\Delta_{\psi} =  \frac{1}{2}}  .
\end{equation}

 To solve for  $a_{{\cal O}_-[n, l]}$ and $a_{{\cal O}_+[n, l]}$ for $n\geq 1$ we need the correlator $g_3(\tau, \vec x)$. 
\subsubsection*{$g_3(\tau, x)$}

From the MFT correlator, we see that 
\begin{eqnarray} \label{g3mft}
g_3(\tau, \vec x) &=& \langle \partial_\mu \psi_\alpha (x)\,  \psi^\dagger_{\beta}  (0) \rangle \gamma^{\mu}_{\beta\alpha} , \\ \nonumber
&=& \sum_{m\in\mathbb{Z}}(-1)^m\frac{d-2\Delta_\psi-1}{[(\tau+m)^2+\vec{x}^2]^{\Delta_\psi+\frac{1}{2}}}.
\end{eqnarray}
We can again perform the expansion in small $x$ in terms of Gegenbauer polynomials as done for the 
correlator $g_2(\tau, \vec x)$. 
This results in 
\begin{align}\label{g3 expans}
	g_3(\tau, \vec x) & =(d-2\Delta_\psi-1) 
	\bigg(\frac{1}{|x|^{ 2\Delta_\psi	+1}}\nonumber\\
	&+\sum_{n=0}^\infty\sum_{l=0,2,\cdots}\frac{2\zeta(2\Delta_\psi+1+2 n+l)(l+\nu)( \Delta_\psi+\frac{1}{2})_{l+n}( \Delta_\psi+\frac{1}{2}-\nu)_n}{n!(\nu)_{l+n+1}} \nonumber\\
	&\qquad\qquad\qquad\qquad\times (2^{-2\Delta_\psi-2 n-l}-1)C_l^{\nu}(\eta)|x|^{2 n+l}\bigg).
\end{align}
We can rewrite the OPE expansion of the correlator   $g_3(\tau, \vec x)$ given in (\ref{opexp3}) as 
\begin{align}\label{g3 ope}
	&g_3(\tau, \vec x)
	=\sum_{n=0}^{\infty}|x|^{2 n}(1+\frac{2n}{d})a_{\mathcal{O}_-}[n,l=0]C_0^{(\nu)} (\eta) \nonumber\\
	&+\sum_{\substack{n=0\\l=1}}^{\infty}|x|^{2 n+l}\left[
	2 (n+1) a_{\mathcal{O}_+}[n+1,l]+\bigg(l+1+\frac{4 n(l+1)}{(l+2)(l+d)}\bigg)a_{\mathcal{O}_-}[n,l] \right] C_l^{(\nu)}(\eta).
	\end{align}
	To obtain this expansion from (\ref{opexp3}) we replace the number of derivatives $J$ by the appropriate spin $l$ of the 
	operators ${\cal O}_+$,  ${\cal O}_-$  and write their conformal dimensions as in (\ref{dimoperator}).  
	We have also separated out  the $l=0$ contribution. 
	Now comparing (\ref{g3 ope}) and (\ref{g3 expans}), we see that  only even spin operators have non-trivial expectation 
	values.  For $l=0$ we obtain 
\begin{align} 
	a_{\mathcal{O}_-}[n,l=0]&=\frac{2 \nu  (\nu +1) \left(2^{-2 (\Delta +n)}-1\right) \left(\Delta +\frac{1}{2}\right)_n 
	 \left(\Delta -\nu +\frac{1}{2}\right)_n (d-2\Delta_\psi-1) }{n! (\nu +n+1) (\nu )_{n+1}}\nonumber\\
	&\qquad\qquad\qquad \qquad \qquad \qquad\qquad \qquad \times \zeta (2 n+2 \Delta +1) .
\end{align}
From $g_2(\tau, \vec x)$, we had already obtained the expectation value of $a_{\mathcal{O}_-}[n,l=0]$ in (\ref{ominusl0}). Note that 
the above equation is identical to that in (\ref{ominusl0}) which serves as an important consistency check of our methods.
Now for $ l=2,4,\ldots $ and $n =0, 1, \cdots $ we have 
\begin{align} \label{lineq1}
		&2 (n+1) a_{\mathcal{O}_+}[n+1,l]+\bigg(l+1+\frac{4 n(l+1)}{(l+2)(l+d)}\bigg)a_{\mathcal{O}_-}[n,l]\nonumber\\
	&=\frac{2 (-2 \Delta_\psi +2 \nu +1) (l+\nu ) \left(2^{-l-2 (\Delta_\psi +n)}-1\right) \left(\Delta_\psi +\frac{1}{2}\right)_{l+n}  
	\left(\Delta_\psi -\nu +\frac{1}{2}\right)_n}{n! (\nu )_{l+n+1}}\nonumber\\
	&\hspace{7cm}\times\zeta (l+2 n+2 \Delta +1).
\end{align}
From the OPE expansion of $g_2(\tau, \vec x)$ we obtained the linear relation between the expectation values given in 
(\ref{1st eq}). 
By replacing $n\rightarrow n+1$ we arrive at the equation
\begin{align}\label{lineq2}
	&a_{\mathcal{O}_+}[n+1,l]+\frac{2(l+1)}{(l+2)(d+l)}a_{\mathcal{O}_-}[n,l]\nonumber\\
	&=\frac{(l+\nu ) (l+2 n+2) \Gamma (\nu ) (1-2^{-l-2 (\Delta +n)}) \Gamma \left(l+n+\Delta +\frac{1}{2}\right)  \Gamma \left(n+\Delta -\nu +\frac{1}{2}\right)}{\Gamma \left(\Delta +\frac{1}{2}\right) \Gamma (n+2) \Gamma \left(\Delta -\nu -\frac{1}{2}\right) \Gamma (l+n+\nu +2)}\nonumber\\
	&\hspace{7cm}\times\zeta (l+2 n+2 \Delta +1),
\end{align}
for $ l=2,4,\ldots $ and $n =0, 1, \cdots $. 

The equations (\ref{lineq1}) and (\ref{lineq2}) are linear independent equations from which we can solve for the expectation 
values  $a_{{\cal O}_+}[n+1, l], a_{{\cal O}_-}[n, l] $ for $n=0, 1, \cdots$ and $l =2, 4, \cdots $. 
We can write the explicit values of these expectation values but they are not very illustrative. 
For $a_{{\cal O}_+}[n=0, l ],\; l =2, 4, \cdots$, the expression is  given in (\ref{oplusn0}), while for 
$a_{{\cal O}_-}[n, 0 ],\; n=0, 1, 2  \cdots$ the expectation value can be read out form (\ref{ominusl0}).
This completes the analysis of obtaining the one-point functions by directly expanding the two-point functions in small $|x|$.

\subsection{OPE inversion  in $d=2$}

Consider the correlator $g_2(x)$ given in (\ref{g2mft}), its OPE expansion is of the form (\ref{genope}). 
Therefore we can apply the inversion formula to obtain the coefficient 
$\hat a(\Delta, l)$ which contains the information of the one-point functions as residues of the
poles in the complex  $\Delta$-plane. 
Using the co-ordinates $z, \bar z$ defined in (\ref{parem}), the  MFT thermal two-point function can be 
written as
\begin{equation} \label{2d2pt}
g_2(z, \bar z) 
=\frac{1}{|x|^{2 \Delta_\psi}}+\sum_{\substack{m=-\infty\\m\ne0}}^{\infty}\frac{(-1)^m}{[(m-z)(m-\bar z)]^{\Delta_\psi+\frac{1}{2}}}\bigg(-\frac{m}{2}\sqrt{\frac{z}{\bar z}}-\frac{m}{2}\sqrt{\frac{\bar z}{ z}}+\sqrt{z \bar z}\bigg).
\end{equation}
Now on substituting $z = rw$ and $\bar z = rw^{-1}$, we see that  the correlator vanishes in  $w$ plane at large $|w|$
as well as small $|w|$
 whenever
$\Delta_{\psi} \geq 1/2$.  In this domain, there is no contribution from the circle or arcs at infinity in the $w$-plane and 
the entire contribution to $\hat a(\Delta, l)$  arises from the discontinuity across the branch cuts \footnote{In the 
domain $ 0< \Delta_{\psi} <1/2 $ we need to evaluate the contributions from the arcs at infinity for only $l =0$.}. 
Therefore we have 
\begin{align}\label{a D J}
	\hat a(\Delta_{\cal O} ,l)=\frac{(1+(-1)^l)}{2\pi}\int_0^1\frac{d\bar z}{\bar z}\int_1^{1/\bar z} \frac{dz}{z} z^{\Delta_\psi-\bar h}\bar z^{\Delta_\psi-h}\text{Disc} [g_2(z, \bar z) ] .
\end{align}
In this expression, the branch cut of (\ref{2d2pt})   in the $w$-plane from $-m $ to $-\infty $ has already been taken 
care by  symmetry  with the 
branch cut from $+m$ to $\infty$  by the inclusion of $(-1)^l$. Therefore we need to restrict ourselves to the branch cut on the 
positive real axis of the $w$-plane.  It is easy to see that this arises solely due to the following discontinuity
\begin{align} \label{discontinuity}
	\text{Disc}\bigg[\frac{1}{((m-z)(m-\bar z))^{\Delta_\psi+\frac{1}{2}}}\bigg]=\frac{2\sin(\pi(\Delta_\psi+\frac{1}{2}))}{[(z-m)(m-\bar z)]^{\Delta_\psi+\frac{1}{2}}}\theta(z-m).
\end{align}
We are therefore led to perform the integral 
\begin{eqnarray} \label{all int}
	&& \hat a(\Delta_{\cal O} ,l) =\frac{(1+(-1)^l)}{2\pi} \times  \\ \nonumber
	&&  \sum_{m=1}^\infty\int_0^1\frac{d\bar z}{\bar z}\int_m^{1/\bar z} \frac{dz}{z} z^{\Delta_\psi-\bar h}\bar z^{\Delta_\psi-h} {(-1)^m}\bigg(-\frac{m}{2}\sqrt{\frac{\bar z}{z}}-\frac{m}{2}\sqrt{\frac{z}{\bar z}}+\sqrt{z \bar z}\bigg)
	 \frac{2\sin(\pi(\Delta_\psi+\frac{1}{2}))}{[(z-m)(m-\bar z)]^{\Delta_\psi+\frac{1}{2}} },   \\ \nonumber
	 &&\qquad\qquad\qquad  \equiv I_1 + I_2 + I_3 .
\end{eqnarray}
Each of the integrals that occur can be carried out term by term. 
Let us examine the first integral which is given by 
\begin{eqnarray}\label{1st int}
I_1 &=& -\frac{(1 + (-1)^l)}{2\pi} \sin\big(\pi(\Delta_\psi + \frac{1}{2} ) \big) \times  \\ \nonumber
	&&\sum_{m=1}^\infty\int_0^1\frac{d\bar z}{\bar z}\int_m^{1/\bar z} \frac{dz}{z} z^{\Delta_\psi-\bar h-\frac{1}{2}}\bar z^{\Delta_\psi-h+\frac{1}{2}}  \frac{(-1)^m m}{[(z-m)(m-\bar z)]^{\Delta_\psi+\frac{1}{2}}} .
\end{eqnarray}
From the integrand it is easy to see that the poles in the $\Delta_{\cal O}$  plane arise due to the small $\bar z$ regime. 
Since we are only interested in the residues at these poles, we can take the upper limit of the integral over $z$ to 
infinity.
Considering the 2nd term in (\ref{1st int}), we have
\begin{align} \label{int1}
	\int_m^{\infty } \frac{z^{\Delta_\psi -\bar h-\frac{3}{2}}}{(z-m)^{\Delta_\psi +\frac{1}{2}}} \, dz=\frac{\Gamma \left(\frac{1}{2}-\Delta_\psi \right)  \Gamma (\bar h+1)}{\Gamma \left(\bar h-\Delta_\psi +\frac{3}{2}\right)}\,m^{-\bar h-1}.
\end{align}
To perform the integral over  $\bar z$, we first expand in small $\bar z$ and integrate term by term, this leads to 
\begin{equation} \label{int2}
(-1)^{m+1} m \int_0^1d\bar z \frac{ \bar z^{\Delta_\psi -h -\frac{1}{2}} }{(m-\bar z)^{\Delta_{\psi} +\frac{1}{2}}} 
= \sum_{n=0}^\infty \frac{(\Delta_\psi + \frac{1}{2})_n}{\Gamma(n+1) (\Delta_\psi -h +\frac{1}{2} +n)} (-1)^{(m+1)} m ^{ -\Delta_\psi - n + \frac{1}{2} }.
\end{equation}
We can use  (\ref{int1}), (\ref{int2}) to obtain the integral
\begin{eqnarray}
I_1 &=&  
\sum_{n=0}^\infty \frac{ ( \Delta _{\psi} + \frac{1}{2} )_n \Gamma( \bar h +1)   ( 1- 2^{-\Delta_\psi-\bar h +\frac{1}{2} -n  })
\zeta( \Delta_\psi + \frac{1}{2} + \bar h + n ) }
{ \Gamma (\Delta_\psi + \frac{1}{2}) \Gamma( \bar h -\Delta_\psi + \frac{3}{2} ) \Gamma(n+1) (\Delta_{\psi} -h +\frac{1}{2} +n) }, 
\\ \nonumber
&& {\rm for}\qquad \qquad l =0, 2, 4, \cdots, 
\end{eqnarray}
where we have summed over $m$. 
We can use the same methods to obtain the remaining two integrals in (\ref{all int}). 
The second integral is given by 
\begin{eqnarray}
I_2  &=&  
\sum_{n=0}^\infty \frac{ ( \Delta _{\psi} + \frac{1}{2} )_n \Gamma( \bar h )   ( 1- 2^{-\Delta_\psi-\bar h +\frac{3}{2} -n  })
\zeta( \Delta_\psi - \frac{1}{2} + \bar h + n ) }
{ \Gamma (\Delta_\psi + \frac{1}{2}) \Gamma( \bar h -\Delta_\psi +\frac{1}{2} ) \Gamma(n+1) (\Delta_{\psi} -h -\frac{1}{2} +n) }, 
\\ \nonumber
&& {\rm for}\qquad \qquad l =0, 2, 4, \cdots, 
\end{eqnarray}
while the 3rd integral is given by 
\begin{eqnarray}
I_3 &=&  
\sum_{n=0}^\infty \frac{ -2 ( \Delta _{\psi} + \frac{1}{2} )_n \Gamma( \bar h)   ( 1- 2^{-\Delta_\psi-\bar h +\frac{1}{2} -n  })
\zeta( \Delta_\psi + \frac{1}{2} + \bar h + n ) }
{ \Gamma (\Delta_\psi + \frac{1}{2}) \Gamma( \bar h -\Delta_\psi + \frac{1}{2} ) \Gamma(n+1) (\Delta_{\psi} -h +\frac{1}{2} +n) }, 
\\ \nonumber
&& {\rm for}\qquad \qquad l =0, 2, 4, \cdots.
\end{eqnarray}

We examine the residues at the poles in $\Delta_{\cal O}$ plane to obtain the one-point functions. 
Consider the residue for the operator with dimension $\Delta_{\cal O}  = 2\Delta_\psi + l -1$, these poles occur only 
in $I_2$ with $n=0$. We obtain the residue
\begin{eqnarray}
a_{{\cal O}_+}[n=0, l] = \frac{2 (\Delta_\psi + \frac{1}{2})_l  ( 1- 2^{-2\Delta_\psi -l  +2} ) \zeta( 2\Delta_\psi + l -1)}{ \Gamma(l) (\Delta_\psi + l -\frac{1}{2} ) }, \qquad l =2, 4, \cdots.
\end{eqnarray}
Comparing this equation with (\ref{oplusn0}), we see that it agrees precisely
 with the one-point function obtained by the brute force expansion 
of the thermal two-point function.   To do this, we relate the residue to the one-point function 
using (\ref{locpoles}) and 
 the one-point function in arbitrary $d$ to $d=2$ using (\ref{2drelation}).
Let us now look at the residue at $\Delta_{\cal O} = 2 \Delta_\psi  +2n +1$, these poles arise in  all the terms
$I_1, I_2$ and $I_3$.
Adding up the contribution from these terms  and using the OPE expansion in (\ref{g2ope}) and (\ref{locpoles})
to identify the residues we get 
\begin{eqnarray}
a_{{\cal O}_-}[n, l=0] = -\frac{ 2 [ (\Delta_\psi + \frac{1}{2})_n ]^2 ( 1- 2\Delta_\psi) ( 1- 2^{ -2(\Delta_\psi +n) } ) 
\zeta(2 \Delta_\psi +2n +1) }{ n! (n+1)! }.
\end{eqnarray}
Again this agrees with the one-point function obtained by the small $x$ expansion in (\ref{ominusl0}) after using 
the relation to $d=2$ in  (\ref{2drelation}). 
Finally let us examine the residues at 
$\Delta_{\cal O} = 2\Delta_{\psi} + 2n +l$ with $l =2, 4, \cdots$ and $n =0, 1, 2, \cdots$. 
Again these poles arise from all terms $I_1, I_2, I_3$, the contribution from $I_2$ can be isolated easily once 
one makes a shift $n\rightarrow n+1$ in $I_2$. 
Summing these residues and again using  (\ref{g2ope}) and (\ref{locpoles}), we obtain the relation,
\begin{align}
	&a_{\mathcal{O}_+}[n+1,l]+\frac{2(l+1)}{(l+2)^2}a_{\mathcal{O}_-}[n,l]\nonumber\\
	&=\frac{(2 \Delta_\psi -1) (l+2 n+2) (1-2^{-l-2 (\Delta_\psi +n)})  \left(\Delta_\psi +\frac{1}{2}\right)_n \left(\Delta_\psi +\frac{1}{2}\right)_{l+n} \zeta (l+2 n+2 \Delta_\psi +1)}{ \Gamma (n+2) \Gamma (l+n+2)}.
\end{align}
We see that this equation coincides precisely with (\ref{lineq2}) once the $d=2$ limit is taken and we use the relation 
(\ref{2drelation}).

Let us now consider the correlator $g_3(x)$, we can write its OPE expansion given in (\ref{g3 ope}) as,
\begin{align}\label{g3}
	g_3(x)=\sum_{n,l=0}^\infty c[n,l]\,|x|^{2 n+l} C_l^{(\nu)}(\eta), 
\end{align}
with
\begin{align} \label{cg3}
	c[n,l=0]&=(1+\frac{2n}{d})a_{\mathcal{O}_-}[n,l=0], \nonumber\\
	c[n,l\ge1]&=\bigg(2 (n+1) a_{\mathcal{O}_+}[n+1,l]+\bigg(l+1+\frac{4 n(l+1)}{(l+2)(l+d)}\bigg)a_{\mathcal{O}_-}[n,l] \bigg).
\end{align}
Now let us examine the MFT thermal correlator $g_3(x)$ which is given by, 
\begin{eqnarray} \label{g3mft2}
g_3(\tau, \vec x) &=& \sum_{m\in\mathbb{Z}}(-1)^m\frac{1-2\Delta_\psi}{[(\tau+m)^2+\vec{x}^2]^{\Delta_\psi+\frac{1}{2}}}.
\end{eqnarray}
This correlator is  very similar to the one studied in \cite{Iliesiu:2018fao}  in MFT for bosonic operators of dimensions $\Delta_\psi$
To obtain the two-point function  (\ref{g3mft2}) from a bosonic MFT correlator, we need to shift 
 $\Delta_\psi \rightarrow \Delta_\psi +\frac{1}{2}$, there is also an insertion of $(-1)^m$ with an overall factor $1-2\Delta_\psi$. 
Taking these changes into account and applying the inversion formula we obtain 
\begin{eqnarray}
	c[n,l]&=&4 (d-2 \Delta_\psi-1) \zeta(2 \Delta_\psi+1+l+2n)\frac{(\Delta_\psi+\frac{1}{2})_{n+l}(\Delta_\psi+\frac{1}{2})_n}{n! \Gamma(n+l+1)}	(2^{-2\Delta_\psi-l-2 n}-1),  \nonumber \\
	&& l =0, 2, 4, \cdots
\end{eqnarray}
We can now use (\ref{cg3}), to identify the one point functions of interest, for $l =0$, we obtain 
\begin{align}
	a_{\mathcal{O}_-}[n,l=0]=\frac{4 (1-2 \Delta_\psi ) \left(2^{-2 (\Delta_\psi +n)}-1\right) \left(\left(\Delta_\psi +\frac{1}{2}\right)_n\right){}^2 \zeta (2 n+2 \Delta_\psi +1)}{(n+1) \Gamma (n+1)^2}.
\end{align}
We see that it agrees with the small $|x|$ expansion of the correlator in (\ref{ominusl0}) after using 
the relation to $d=2$ in  (\ref{2drelation}). 
For $l\geq 2$ we obtain  the relation
\begin{align}
	&2 (n+1) a_{\mathcal{O}_+}[n+1,l]+\bigg(l+1+\frac{4 n(l+1)}{(l+2)^2}\bigg)a_{\mathcal{O}_-}[n,l]\nonumber\\
	&=4 (1-2 \Delta_\psi) \zeta(2 \Delta_\psi+1+l+2n)\frac{(\Delta_\psi+\frac{1}{2})_{n+l}(\Delta_\psi+\frac{1}{2})_n}{n! \Gamma(n+l+1)}	(2^{-2\Delta_\psi-l-2 n}-1).
\end{align}
Again this equation precisely agrees with the second  linear equation (\ref{lineq2}) between one point functions of ${\cal O}_{-}$, 
${\cal O}_+$  obtained using the  brute force expansion in small $|x|$, 
when $d=2$ and after using the relation (\ref{2drelation}).

\subsection{OPE inversion   in  $d>2$}

 The inversion formula for  $d>2$ dimensions is given in (\ref{finalinversion}). 
 \begin{eqnarray}\label{finalinversion1}
\hat a( \Delta_{} , l) &=&  \hat a_{\rm disc} ( \Delta_{}, l) + \theta( l_0 -l)  \hat a_{\rm arcs} ( \Delta_{} -l), \\ \nonumber
\end{eqnarray}
The correlator of interest is the mean field theory correlator  $g_2(x)$ given in (\ref{2d2pt}), the form of the MFT correlator is 
invariant across dimensions once we choose the kinematics as discussed around  (\ref{parem}). 
The contribution from the $\hat a_{\rm arcs}$ at infinity is given by  (\ref{db2arc})
\begin{eqnarray}\label{db2arc1}
\hat a_{\rm arcs} (\Delta_{},  l)  &=& 2 K_l  \int_0^1 \frac{dr}{r^{\Delta_{}  +1 - 2\Delta_\psi} }  \times  \\ \nonumber
&& \oint \frac{dw}{i w} \lim_{|w| \rightarrow \infty} 
\left[ (  w - w^{-1} )^{2\nu} 
 F_l(w^{-1}) e^{i\pi\nu} g( r, w) 
\right].
\end{eqnarray}
Using the definition of $F_l(w)$ given in (\ref{genhyp1}), we see that the contribution reduces to 
\begin{equation} \label{contriarc}
\hat a_{\rm arcs} (\Delta_{},  l)  =  2 K_l  \int_0^1 \frac{dr}{r^{\Delta_{}  +1 - 2\Delta_\psi} }  \times 
 \oint \frac{dw}{i w} \lim_{|w| \rightarrow \infty}  w^{-l} g_2(r, w).
\end{equation}
From the expression of the 2-point function for $g_2$ given in (\ref{2d2pt}),   it vanishes as $|w|^{-\Delta +\frac{1}{2}}$ for 
large $|w|$. 
Therefore again as in the case of $d=2$,  if  $\Delta_\psi \geq \frac{1}{2}$, we see that  there is no contribution 
from the arcs at infinity and for $0 <\Delta_\psi <\frac{1}{2} $, we just need to include the arc contribution for 
the $l=0$ case.  As in  $d=2$  dimensions, we will take $\Delta_\psi\geq \frac{1}{2}$. 
We are thus led to evaluating only the contribution from the discontinuity across the 
 branch cuts, which is given by 
\begin{eqnarray}
\hat a_{\rm disc} ( \Delta_{}, l)&=& ( 1 + (-1)^l) K_l
 \int_0^1\frac{d\bar z}{\bar z} \int_1^{ \frac{1}{\bar z} } \frac{dz}{z} (z\bar z)^{\Delta_{\psi} -\frac{\Delta_{}}{2} -\nu}
F_l \Big( \sqrt{\frac{\bar z}{z}} \Big) {\rm Disc} [g(z, \bar z) ], \\ \nonumber
K_l &=& \frac{\Gamma(l+1) \Gamma(\nu) }{ 4\pi \Gamma(l+\nu) }.
\end{eqnarray}
The discontinuity across the branch cuts is given in (\ref{discontinuity}). 
Substituting this, we obtain the following expression for the one point function
\begin{eqnarray}\label{ad_2}
	\hat a(\Delta, l) &=&(1+(-1)^l) K_l\sum_{m=1}^\infty \int_{0}^{1}\frac{d\bar z}{\bar z}\int_{1}^{\text{max}(m,1/\bar z)}\frac{dz}{z} (z \bar z)^{\Delta_\psi-\frac{\Delta}{2}-\nu} (z-\bar z)^{2\nu} F_J\bigg(\sqrt{\frac{\bar z}{z}}\bigg) \nonumber\\
	&&\qquad\qquad\qquad\times\frac{(-1)^m2\sin(\pi(\Delta_\psi+\frac{1}{2}))}{[(z-m)(m-\bar z)]^{\Delta_\psi+\frac{1}{2}}} \bigg(-\frac{m}{2}\sqrt{\frac{z}{\bar z}}-\frac{m}{2}\sqrt{\frac{\bar z}{ z}}+\sqrt{z \bar z}\bigg) \nonumber ,\\
	&& \qquad\qquad\qquad \equiv I_1 + I_2 +I_3.
\end{eqnarray}
The last line defines the three integrals which must be done to obtain $\hat a(\Delta, l)$. 
Let us  examine the first integral 
\begin{align}
	I_1& =2 K_l \sin\big(\pi( \Delta_\psi + \frac{1}{2} ) \big) \\ \nonumber
	 & \times \int_0^1\frac{d\bar z}{\bar z}\int_m^{{\rm max} (m, \frac{1}{\bar z} )}
	   \frac{dz}{z} \frac{(-1)^m (-m)  F_J\left(\sqrt{\frac{\bar z}{z}}\right) (z-\bar z)^{2 \nu } (z \bar z)^{-\frac{\Delta }{2}+\Delta_\psi-\nu }}{ (z-m)^{\Delta_\psi+\frac{1}{2}} (m-\bar z)^{\Delta_\psi+\frac{1}{2}}}\sqrt{\frac{z}{\bar z}}, \\ \nonumber
	   & l = 0, 2, \cdots.
\end{align}
It is convenient to define new  variables 
\begin{equation} \label{changver}
\bar z = m z'  \bar z', \qquad z = m z'.
\end{equation}
Using the $(z', \bar z')$ variables we obtain the integral
\begin{align}
I_1(m)  &=
	 2 m^{-\Delta }  (-1)^{m+1} K_l \sin\big(\pi( \Delta_\psi + \frac{1}{2} ) \big)  
	 \int_0^1 \frac{d\bar z}{\bar z}
	\int_1^{{\rm max} ( \frac{1}{m{\sqrt{\bar z}}}, 1) } \frac{dz}{z}   \Bigg[ \\ \nonumber
	 &   (z-1)^{-\Delta_\psi-\frac{1}{2}} (1-z \bar z)^{-\Delta_\psi-\frac{1}{2}} (z-z \bar z)^{2 \nu } \left(z^2 \bar z\right)^{-\frac{\Delta }{2}+\Delta_\psi-\nu }\bar z ^{-\frac{1}{2}} F_J\left(\sqrt{\bar z}\right) \Bigg].
\end{align}
Here we have re-named the primed variables and isolated the $m$-th term in the sum. 
As in the case of $d=2$ and in \cite{Iliesiu:2018fao},  
we expect the poles to arise at 
$\bar z \rightarrow 0$ in the integrand. 
Therefore, we can take the upper limit of the $z$ integration to $\infty$ and the integrals over $z$ and $\bar z$ factorize.
Performing the integral over $z$ from $1$ to  $\infty$ we obtain 
\begin{align}
	I_1(m) &= 
	\frac{ 2\pi  K_l m^{-\Delta } (-1)^{m+1} \Gamma (\Delta +1)  }{ 
	\Gamma( \frac{1}{2} + \Delta_\psi) \Gamma \left(\Delta -\Delta_\psi+\frac{3}{2}\right)} 
	 \int_{0}^{1}d\bar z\sqrt{\frac{1}{\bar z}}\, (-\bar z)^{-\Delta_\psi-\frac{1}{2}} (1-\bar z)^{2 \nu }  \bar z^{-\frac{\Delta }{2}+\Delta_\psi-\nu -1} \nonumber \\ 
	 &  \qquad\qquad\qquad\qquad\qquad
	\times\, _2F_1\left(\Delta +1,\Delta_\psi+\frac{1}{2};\Delta -\Delta_\psi+\frac{3}{2};\frac{1}{\bar z}\right)F_J\left(\sqrt{\bar z}\right).
\end{align}
We can now expand in small $\bar z$ and perform the integrals term by term.  The leading and the sub-leading terms are given by 
\begin{eqnarray}
I_1(m) &=& \frac{2\pi K_l  m^{-\Delta} (-1)^{m+1}\Gamma( \Delta-\Delta_\psi + \frac{1}{2} ) }
{\Gamma( \frac{1}{2} + \Delta_\psi)  \Gamma( 1+ \Delta - 2\Delta_\psi)}  \int_0^1 d\bar z
\bar z^{( -\frac{\Delta}{2} + \frac{l}{2} +\Delta_\psi -\frac{3}{2} ) } \\ \nonumber
&& \times  \left[ 1+   \bar z \Big( \frac{\nu ( l + 2 \nu)}{l+ \nu +1} +
 \frac{(\Delta -2\Delta_\psi)( 1+ 2\Delta_\psi)}{2\Delta - 2\Delta_\psi -1}   \Big) + O(\bar z^2) \right].
\end{eqnarray}
Performing the sum over $m$ and the integral over $\bar z$, we obtain 
\begin{eqnarray} \label{I1mft}
&& I_1 = \frac{\Gamma(\nu) \Gamma( l +1)   \Gamma( \Delta-\Delta_\psi + \frac{1}{2} ) ( 1- 2^{ 1- \Delta}) \zeta( \Delta) }{
\Gamma( l+\nu) \Gamma( \frac{1}{2} + \Delta_\psi) \Gamma( 1+ \Delta - 2\Delta_\psi) } \\ \nonumber
&& \times \left[ \frac{1}{( - \Delta + l + 2\Delta_\psi -1) } +  \Big( \frac{\nu ( l + 2 \nu)}{l+ \nu +1} +
 \frac{(\Delta -2\Delta_\psi)( 1+ 2\Delta_\psi)}{2\Delta - 2\Delta_\psi -1}  -2 \nu  \Big) \frac{1} {( - \Delta + l + 2\Delta_\psi +1) } 
  \right] \\ \nonumber
 && \qquad\qquad\qquad +\cdots.
\end{eqnarray}
Let us evaluate the  residue at the pole $\Delta = 2\Delta_\psi +l - 1$, from the general OPE expansion 
in  (\ref{g2ope}) we see residue results in 
 following one point function 
\begin{eqnarray} \label{opexpmft}
a_{{\cal O}_+}[n =0, l ] = - \frac{l   (\Delta_\psi + \frac{1}{2})_l  ( 1- 2^{ - \Delta_\psi-l+2}) \zeta( 2\Delta_\psi + l -1)    }
{ (\nu)_l ( \Delta_\psi + l -\frac{1}{2} )  }.
\end{eqnarray}
This result precisely coincides with the one point function obtained in (\ref{oplusn0}) by the directly expanding the correlator 
$g_2(x)$ in the small $x$ expansion. 

We can proceed on similar lines and evaluate the leading contribution of $I_2$ in the small $\bar z$ expansion of the 
integrand. This  allows us to obtain the 
residues at $\Delta = 2\Delta_\psi +l +1$.  These are given by 
\begin{eqnarray}\label{I2mft}
I_2 =  \frac{\Gamma(\nu) \Gamma( l +1)   \Gamma( \Delta-\Delta_\psi + \frac{1}{2} ) ( 1- 2^{ 1- \Delta}) \zeta( \Delta) }{
\Gamma( l+\nu) \Gamma( \frac{1}{2} + \Delta_\psi) \Gamma( 1+ \Delta - 2\Delta_\psi)  ( - \Delta + l + 2\Delta_\psi +1) } 
+ \cdots ,
\end{eqnarray}
and 
\begin{eqnarray} \label{I3mft}
I_3 =- 2  \frac{\Gamma(\nu) \Gamma( l +1)   \Gamma( \Delta-\Delta_\psi - \frac{1}{2} ) ( 1- 2^{ 1- \Delta}) \zeta( \Delta) }{
\Gamma( l+\nu) \Gamma( \frac{1}{2} + \Delta_\psi) \Gamma(  \Delta - 2\Delta_\psi)  ( - \Delta + l + 2\Delta_\psi +1) } 
+ \cdots .
\end{eqnarray}

From the expression of $I_1$ in (\ref{I1mft}) we see that too contains a pole at $\Delta = 2\Delta_\psi +l +1$. 
Adding up the residue at $\Delta = 2\Delta_\psi +1, l =0$ allows us to obtain the one point function 
\begin{equation}
a_{{\cal O}_-}[n=0, l =0] = 2 ( 1- 2^{-2\Delta_\psi} ) ( 2\Delta_\psi - 2\nu -1) \zeta( 2\Delta_\psi +1) .
\end{equation}
Here we have related the residue to the one point function using (\ref{g2ope}). 
We see that the result is in precise agreement with (\ref{ominusl0}) which is obtained by the brute force expansion of the 
correlator $g_2(x)$. 
Finally for $l>0$ we can combine the all the residues  at $\Delta = 2\Delta_\psi +l +1$
 from $I_1, I_2, I_3$  from (\ref{I1mft}), (\ref{I2mft}), (\ref{I3mft}) to obtain the equation 
\begin{eqnarray}
&& a_{{\rm O}_+}[1, l ] + \frac{2 (l +1)}{ (l +2) ( d +l )} a_{{\cal O}_-}[0, l ]  = \\ \nonumber
&& \qquad\qquad\qquad \frac{ (l +2) (\Delta_\psi + \frac{1}{2})_l ( \Delta_\psi - \nu -\frac{1}{2}) ( 1- 2^{-\Delta_\psi -l}) 
\zeta( 2\Delta_\psi + l+1)}{(\nu)_l ( l +\nu +1) }.
\end{eqnarray}
Again we have used the OPE expansion (\ref{g2ope}) to identify the linear combination of the one-point functions with the 
residue.   The result  precisely coincides with  the equation (\ref{lineq2}) obtained by the brute  force expansion 
of $g(x)$ in small $|x|$.

Let us finally examine the correlator $g_3(x)$ given in (\ref{g3mft}). 
As we have discussed for the case of $d=2$, this two-point function is similar to the two-point function of scalars 
in MFT
studied in \cite{Iliesiu:2018fao} but with a  $(-1)^m$ inserted in the sum over images. We also need to replace
 $\Delta_\psi \rightarrow \Delta_\psi +\frac{1}{2}$ and multiply by the overall factor $2\nu +1 -2\Delta_\psi$.
Consider the expansion of $g_3(x)$ given in (\ref{g3}), using the inversion formula given in \cite{Iliesiu:2018fao} together with the 
modifications mentioned above, we obtain  the residue at $\Delta = 2\Delta_\psi +1  +l $
\footnote{This is equation (4.19) of \cite{Iliesiu:2018fao}.},
\begin{eqnarray}
c(0, l) = 2 \frac{ ( \Delta_\psi + \frac{1}{2} )_l }{( \nu)_l }
( 2\Delta_\psi - 2\nu -1) ( 1 - 2^{-(2\Delta_\psi +l) }) \zeta(  2\Delta_\psi +1  +l ).
\end{eqnarray}
We can now use the equations (\ref{cg3}) to identify the one point functions. 
We obtain for $l=0$,
\begin{eqnarray}
c(0, 0) = a_{{\cal O}_-}[n=0, l =0] = 2 ( 1- 2^{-2\Delta_\psi} ) ( 2\Delta_\psi - 2\nu -1) \zeta( 2\Delta_\psi +1) .
\end{eqnarray}
For $l =2,  4, \cdots$. we get 
\begin{eqnarray}
c(0, l\geq 2) &=&  2 a_{{\cal O}_+} [1, l ]  + ( l +1) a_{{\cal O}_-}[0, l]  \\ \nonumber
&=&
2 \frac{ ( \Delta_\psi + \frac{1}{2} )_l }{( \nu)_l }
( 2\Delta_\psi - 2\nu -1) ( 1 - 2^{-(2\Delta_\psi +l) }) \zeta(  2\Delta_\psi +1  +l ).
\end{eqnarray}
Note that this equation precisely coincides with the linear equation relating these one 
 point functions given in (\ref{lineq1}) obtained by the small $x$ expansion of the correlator $g_3(x)$. 
 
 This concludes the discussion of using the inversion formula for the MFT of fermions. It is important to mention that
 the application of the inversion formula on the correlator $g_2(x)$ is not related to the one studied for the 
 scalars in \cite{Iliesiu:2018fao}.   The  MFT correlator $g_2(x)$  in (\ref{2d2pt})  has factors which involve $\sqrt{z}, \sqrt{\bar z}$, 
 inspite of this, the location of  branch cuts in the complex $w$-plane  falls into the general discussion of \cite{Iliesiu:2018fao}.
 Therefore we could apply the inversion formula leading to results which agree with the brute force small $x$ 
 expansion of the correlator $g_2(x)$.

\section{The critical Gross-Neveu model at large $N$} \label{secgnmodel}

In this section we study thermal one-point functions in the critical $U(N)$ Gross-Neveu model at large $N$ in arbitrary 
odd $d = 2k+1$ dimensions. 
The theory is defined using the action 
\begin{equation}
S = \int d^{d} x\left[  i\bar\psi_a  \gamma^\mu \partial_\mu \psi_a+ \frac{\lambda}{N} ( \bar \psi_a \psi_a)^2 
\right].
\end{equation}
where $N$ is the number of fermions and $a=1, 2,\, \cdots, N$.  For $d>3$, the interaction is non-renormalizable, 
nevertheless in \cite{Filothodoros:2018pdj,Petkou:2018ynm,Filothodoros:2023ppi}, it has been argued that on choosing a definite prescription to evaluate the partition function
one is led to a gap equation which is independent of the cutoff at large $N$ \footnote{A similar procedure has been followed 
to obtain the gap equation and the thermal mass  for the bosonic $O(N)$ model in arbitrary $d$ dimensions 
\cite{Petkou:2018ynm,Giombi:2019upv}.}. 
The derivation of the gap equation is given in the appendix \ref{appendgap}, the equation is given by 
\begin{eqnarray} \label{gngap}
2 ( 2m_{\rm th})^k \sum_{n=0}^{k-1} \frac{ ( k-n)_{2n} }{ ( 2m_{\rm th})^n n! }  {\rm Li}_{k+n} ( - e^{-m_{\rm th} })
+ \frac{(m_{\rm th}) ^{2k}  \sqrt{\pi}}{\Gamma( \frac{1}{2} + k) \cos\pi k} =0.
\end{eqnarray}
  The gap equation has a real solution for the 
thermal mass $m_{\rm th}$ in dimensions 
 $d=2k +1$, with $k =2, 4, 6, \cdots$. While
for $k =1, 3, 5, \cdots$, the gap equation has complex solutions for $m_{\rm th}$.  
The table \ref{table} lists the thermal masses of the Gross-Neveu model in  various dimensions.

In \cite{Iliesiu:2018fao}, it was observed that for the bosonic $O(N)$ model in $d=3$  the corresponding  gap equation can be 
obtained  by demanding the  scalar  bilinear 
$\phi^a\phi_a$ with dimension $\Delta =1$ does not exist in the spectrum at the critical point. 
This observation was also  seen to hold true for the $O(N)$ model in arbitrary odd dimensions in \cite{Petkou:2018ynm}.
This paper also studied the application of the inversion formula for the correlator $g_1(x)$  in 
Gross-Neveu model  in arbitrary odd dimensions. 
Similar to the bosonic case, it was observed that the gap equation of the Gross-Neveu model 
 in (\ref{gngap}) can be obtained by demanding that the 
operator ${\cal O}_0[0,0]$ or schematically the bilinear 
$\bar\psi \psi$ with dimensions $\Delta = 2 k$ does not exist in the spectrum at the critical point.

In this section, we  
study the correlator $g_2(x)$ in detail and also examine the correlator $g_1(x)$ and $g_3(x)$. 
We see that the  absence of the operator ${\cal O}_-[0,0]$  which is 
schematically of the form $\bar \psi \gamma^\mu \partial _\mu \psi$ in the  spectrum 
also leads to the identical gap equation in (\ref{gngap}).  
We then evaluate the one-point functions of operators ${\cal O}_{+}[0,l]$ or operators of the  form 
\begin{equation}
{\cal O}_{+}[0, l]: \quad  \bar\psi \gamma_{\mu_1}\partial_{\mu_2} \cdots \partial_{\mu_l} \psi,  \qquad \Delta =  2k + l-1 .
\end{equation}
We show that the one-point functions of these operators are given by 
\begin{eqnarray} \label{gnao+}
a_{{\cal O}_+}[0, l]  &=&  \frac{l \, (m_{\rm th})^{l +k-1}}{ \pi^k 2^{l+k}  ( k- \frac{1}{2})_l}
\sum_{n=0}^{l +k -1}  \frac{ (k +l -n)_{2n}  }{ 2^n n! (m_{\rm th})^{n}}  {\rm Li}_{k+n} ( - e^{ -m_{\rm th}} ), \\ \nonumber
&& l = 2, 4, \cdots.
\end{eqnarray}
The one-point function $a_{{\cal O}_+}[0, l=2]$ corresponds to the stress tensor.
In the appendix \ref{appendgap}, we have evaluated the stress tensor directly from the partition function. 
This is given by 
\begin{eqnarray} \label{stresspart}
	T_{00} &=& \frac{ (m_{\rm th})^{2k+1}}{ 2^{k+2}  \pi^{k-\frac{1}{2} } \Gamma( k + \frac{3}{2} ) \cos \pi k } \\ \nonumber
	&& 
	+ \frac{ (m_{\rm th})^{k+1} }{\pi^k} \sum_{n=0}^{k+1} \frac{ [ ( k+n)^2 + (k-n)] ( k-n+2)_{2n-2} }{ 2^n n! (m_{\rm th})^n}
	{\rm Li}_{k+n} ( - e^{-m_{\rm th}}).
\end{eqnarray}
Manifestly the one-point function 
  $a_{{\cal O}_+}[0, l=2]$ does not seem to agree with the  stress tensor  in (\ref{stresspart}), however on substituting 
  the value of $m_{\rm th}$  from the gap equation (\ref{gngap})
  and  scaling by the overall dimension dependent constant they precisely coincide. 
  The overall scaling is because  the one-point function $a_{{\cal O}_+}[0, l=2]$ also contains the structure constant
  $f_{\psi^\dagger\psi T}$ and the normalisation of the two point function $c_T$.
  The table \ref{table}  also compares the  values of  $a_{{\cal O}_+}[0, l=2]$ with the stress tensor.

\begin{table}[ht]
	\begin{center}
		\begin{tabular}{ |c| c| c| c| c|}\hline

			$d$& $m_{th}(\in \mathbb{R})$&$a_{\mathcal{O}_+}[0,l]$&$T_{00}$& $a_T= 2^k k (2 k-1) a_{\mathcal{O}_+}[0,l]$\\ \hline
			
			5 & 1.48051 & $-0.04048 $&$ -0.971519$ & $-0.971519$ \\
			9 & 2.86003 & $-0.0125136$ & $-5.6061 $& $-5.6061$ \\
			13 & 4.24178 & $-0.0159513$ & $-67.3783$ & $-67.3783$ \\
			17 & 5.6273 & $-0.0460522$ & $-1414.72$ & $-1414.72$ \\
			21 & 7.01451 & $-0.23691$ & $-46093.1$ & $-46093.1$ \\
			25 & 8.40257 & $-1.90627$ & $-2.15503\times 10^6 $& $-2.15503\times 10^6$ \\
			29 & 9.7911 & $-22.0996$ & $-1.36866\times 10^8$ & $-1.36866\times 10^8 $\\
			33 & 11.1799 & $-348.833$ & $-1.13391\times 10^{10} $& $-1.13391\times 10^{10}$ \\
			37 & 12.569 & $-7193.26$ & $-1.18797\times 10^{12}$ & $-1.18797\times 10^{12}$ \\
			41 & 13.9581 & $-187758.14$ & $-1.53565\times 10^{14}$ & $-1.53565\times 10^{14}$ \\
			45 & 15.3474 & $-6.05104\times 10^6$ & $-2.40094\times 10^{16}$ & $-2.40094\times 10^{16}$ \\
			49 & 16.7368 & $-2.35981\times 10^8 $& $-4.46587\times 10^{18}$ & $-4.46587\times 10^{18}$ \\
			53 & 18.1262 & $-1.09528\times 10^{10}$ & $-9.74651\times 10^{20}$ & $-9.74651\times 10^{20} $\\
			57 & 19.5156 & $-5.96646\times 10^{11}$ & $-2.46648\times 10^{23}$ & $-2.46648\times 10^{23}$ \\
			61 & 20.9051 & $-3.76957\times 10^{13}$ & $-7.16416\times 10^{25}$ & $-7.16416\times 10^{25}$ \\
			65 & 22.2946 & $-2.73404\times 10^{15}$ & $-2.36731\times 10^{28}$ & $-2.36731\times 10^{28}$ \\
			69 & 23.6842 & $-2.25624\times 10^{17}$ & $-8.82995\times 10^{30}$ & $-8.82995\times 10^{30}$ \\
			73 & 25.0737 & $-2.10199\times 10^{19}$ & $-3.69208\times 10^{33}$ & $-3.69208\times 10^{33}$ \\
			77 & 26.4633 & $-2.19548\times 10^{21}$ & $-1.71994\times 10^{36}$ & $-1.71994\times 10^{36}$ \\
			81 & 27.8529 & $-2.55503\times 10^{23}$ & $-8.87735\times 10^{38}$ & $-8.87735\times 10^{38}$ \\

\hline

	\end{tabular}\end{center}
\caption{The table shows the agreement in the value of stress tensor evaluated from the partition function   given in 
(\ref{stresspart} ) with the one-point function from
OPE inversion formula given in (\ref{gnao+}) at  $l=2$. We need the $2^k$, the dimension of the Dirac spinor in the last column,  since 
we had factored this out in the two point function (\ref{mth2pt}), the rest of the factors in  the last column  are because of the 
presence of the structure constant $f_{\psi^\dagger\psi {\cal O}}$
 and the normalization of the 2-pt functions $c_{\cal O}$ in $a_{{\cal O}_+}[0,2]$. }\label{table}
\end{table}

Our analysis also shows that 
the  one-point functions  in (\ref{gnao+}) are related  to that of  one-point functions of 
${\cal O}_0[0, l]$   or operators of the form
\begin{equation}
{\cal O}_{0}[0, l]: \quad \bar\psi \partial_{\mu_1}\partial_{\mu_2} \cdots \partial_{\mu_l} \psi  \qquad \Delta_{\cal O} =  2k + l.
\end{equation}
The relation is given by 
\begin{equation}
a_{{\cal O}_0}[0, l] = m_{th} a_{{\cal O}_+}[0, l ] .
\end{equation}
Similarly evaluating the one-point functions 
$a_{{\cal O}_+}[1, l]$ and $a_{{\cal O}_-}[0,l]$, we see that they are related to the one-point functions 
$a_{{\cal O}_0}[0, l]$ by factors which depend on $l$ and $m_{\rm th}$. 
Thus by explicit calculation, we see that the one-point functions of all fermion bi-linears are related to $a_{{\cal O}_0}[0, l]$ 
which occurs in the OPE expansion of $g_1(x)$. 
It is interesting to contrast this with MFT, in which $g_1(x)$ trivially vanished and did not contain any 
one-point functions.

\subsection{OPE inversion on $g_2(x)$}

The two-point function of fermions with a thermal mass $m_{\rm th}$ and at finite temperature is given by 
\begin{align} \label{mth2pt}
	\langle \psi_\alpha (x)  \psi_\beta^\dagger (0)\rangle = \frac{i}{2^{\frac{d-1}{2}}}\sum_{n,\, k_0 = 2\pi ( n +\frac{1}{2} ) }\int \frac{d^{d-1}k}{(2\pi)^{d-1}}\frac{\gamma^\mu_{\alpha\beta} k_\mu -i m_{th}\delta_{\alpha\beta}}{k^2+m_{th}^2} e^{ikx}.
\end{align}
Note that this correlator is anti-periodic under the shift $\tau \rightarrow \tau +1$, again we have divided 
by the dimension of the Dirac spinor so that traces over $\gamma$ matrices gives unity. 
It is useful for us to write down the correlator with $\bar\psi$ and $\psi$ inter-changed for the constructions
of $g_2(x)$ and $g_3(x)$. From (\ref{mth2pt})  we obtain
\begin{equation} \label{gnftcor1}
	\langle \psi^\dagger_\beta (x)  \psi_\alpha (0)\rangle = \frac{i}{2^{\frac{d-1}{2}}}\sum_{n,\, k_0 = 2\pi ( n +\frac{1}{2} ) }\int \frac{d^{d-1}k}{(2\pi)^{d-1}}\frac{\gamma^\mu_{\alpha\beta} k_\mu +i m_{th}\delta_{\alpha\beta}}{k^2+m_{th}^2} e^{ikx}.
\end{equation}
To derive (\ref{gnftcor1}) from (\ref{mth2pt}), we inter-change the fermions,  change the dummy variables of integrations and 
summations and also use translation invariance. 
Using this  we can write down the correlator $g_2(x)$ using the definition in (\ref{corr})
	\begin{align}\label{2ptfnv}
	g_2(\tau, \vec x)  =\frac{i}{|x|}&\sum_{n,\, k_0 = 2\pi ( n +\frac{1}{2} )}\int \frac{d^{d-1}k}{(2\pi)^{d-1}}\frac{k_\mu x^\mu}{k^2+m_{th}^2} e^{ikx}, \nonumber\\
	=&\frac{x^\mu \partial_\mu  }{|x|}\sum_{n, \, k_0 = 2\pi ( n +\frac{1}{2} )}\int \frac{d^{d-1}k}{(2\pi)^{d-1}}\frac{e^{ikx}}{k^2+m_{th}^2} .
\end{align}
To perform the integral, we first use the Poisson re-summation formula to convert the sum over Matsubara frequencies to 
sum over images in $\tau$. 
\begin{align}
	\sum_{n\in \mathbb{Z}}f\big[(2n+1)\pi\big]=\sum_{n\in \mathbb{Z}}(-1)^n \int_{-\infty}^{\infty}\frac{d\omega}{2\pi} f(\omega) e^{in\omega} .
\end{align}
Applying the re-summation on the integral in  (\ref{2ptfnv}) and performing the resultant integral,  we obtain 
\begin{eqnarray}\label{poiresum}
& \sum_{ n,\, k_0 = 2\pi ( n +\frac{1}{2} )}\int \frac{d^{d-1}k}{(2\pi)^{d-1}}\frac{e^{i\vec{k}.\vec{x}}e^{ik_0 \tau}}{ k_0^2 +\vec{k}^2
+ m_{\rm th}^2}
= \sum_{n\in \mathbb{Z}}(-1)^n \int\frac{d^{d-1}k d\omega}{(2\pi)^d}\frac{e^{i\vec{k}\cdot \vec{x}}e^{i\omega(\tau+n)}}{\omega^2+\vec{k}^2+m_{\rm th}^2},  \nonumber \\
&= \sum_{n\in \mathbb{Z}}(-1)^n(2 \pi )^{-\frac{d}{2}} \bigg(\frac{|x^{(n)}|}{m_{\rm th}}\bigg)^{1-\frac{d}{2}} K_{\frac{d}{2} -1}(m_{\rm th} |x^{(n)}|).
\end{eqnarray}
Here $x^{(n)} = (\tau +n, \vec x)$, choose the configuration given in (\ref{parem}) we can express the correlator as
a function of $(z, \bar z)$
\begin{eqnarray} \label{gng2}
g_2(z, \bar z) 
=\sum_{m\in \mathbb{Z}}(-1)^{m+1}\big(\frac{m_{\rm th}}{2\pi} \big)^{\frac{d}{2}} 
  \bigg[-\frac{m}{2} \sqrt{\frac{z}{\bar z}} - \frac{m}{2} \sqrt{\frac{\bar z}{ z}}
+\sqrt{z \bar z}\bigg] 
 \frac{K_{\frac{d}{2}}\left(m_{\rm th} \sqrt{(m-z)(m-\bar z)}\right)}{\big(\sqrt{(m-z)(m-\bar z)}\,\big)^{d/2}}. \nonumber 
\\
\end{eqnarray}
One consistency check of this correlator is the following, on taking the $m_{\rm th}\rightarrow 0$ limit it is proportional
to the MFT correlator in (\ref{g2mft})  or (\ref{2d2pt}) with $\Delta_\psi = k $
\begin{equation}\label{mftgnrel}
g_2(z, \bar z)|_{{\rm Gross-Neveu} ,\; m_{\rm th}\rightarrow 0} =  -\frac{\Gamma( k + \frac{1}{2} )}{2 \pi^{k +\frac{1}{2} }}
g_2(z, \bar z)|_{{\rm MFT}, \; \Delta_{\psi} = k }.
\end{equation}
On comparing the two-point function in (\ref{gng2}), 
 with the corresponding one for the bosonic $O(N)$ model studied in 
\cite{Iliesiu:2018fao},  we have an insertion of $(-1)^m$ since we are dealing with fermions.
We also have the factor in the  square brackets in addition to the Bessel function. 
Examining this factor in the $w$-plane where $w$ is defined as (\ref{parem}), we see that this factor does not 
affect the branch cut structure  present in the Bessel function together with the
 factor  $\big(\sqrt{(m-z)(m-\bar z)}\,\big)^{-d/2}$.
Therefore, the branch cuts in the $w$-plane are as assumed in  section \ref{eucinv} and we can proceed to apply the 
inversion formula. 
Let us first write down the contribution from the discontinuities
\begin{eqnarray} \label{gninverf}
\hat a_{\rm disc} ( \Delta, l ) = 2 K_l \int_0^1 \frac{d\bar z}{\bar z} \int_1^{\frac{1}{\bar z} } \frac{dz}{ z}
( z- \bar z)^{2\nu} ( z\bar z)^{\frac{1}{2} - \frac{\Delta}{2}} F_l \Big( \sqrt{ \frac{\bar z}{z} } \Big) {\rm Disc}[g_2(z, \bar z) ].
\end{eqnarray}
Here we have substituted $\Delta_\psi = \frac{d-1}{2}$ and $l \in 2\mathbb{Z}$. At this point we note that we have taken
some input from the perturbative results of \cite{Moshe:2003xn,Fei:2014yja}, that the fundamental field does not acquire anomalous dimensions 
at  large $N$ for the Gross-Neveu model. 
It should also be noted that the branch cut for each term in (\ref{gng2}) depends on $m$ and the integration range in $z$ depends on 
$m$ for each term. 
As we have mentioned earlier, the branch cut in the $w$-plane can be obtained from the relation
\begin{equation} \label{bcrelation}
{\rm Disc}\left[ \frac{K_{k +\frac{1}{2} } (\sqrt{-x}) }{(-x)^{\frac{k}{2} +\frac{1}{4}}} \right]
= \pi \frac{ J_{-\frac{1}{2} -k} (\sqrt{x})}{ x^{\frac{k}{2} +\frac{1}{4}}} \theta(x) .
\end{equation}
Substituting this relation in (\ref{gninverf}) we obtain 
\begin{eqnarray}
\hat a_{\rm disc} ( \Delta, l ) &=& 2\pi K_l  \bigg(\frac{m_{\rm th}}{2\pi} \bigg)^{\frac{d}{2}} 
\sum_{m=1}^\infty \int_0^1 \frac{d\bar z}{\bar z} \int_1^{{\rm max } (m, \frac{1}{\bar z} )} \frac{dz}{z} ( z- \bar z)^{2\nu} ( z\bar z)^{\frac{1}{2} - \frac{\Delta}{2}} F_l \Big( \sqrt{ \frac{\bar z}{z} } \Big) \\ \nonumber
&&\times  \frac{ (-1)^{m+1}  J_{-\frac{1}{2} -k} ( m_{\rm th}  \sqrt{(z-m)(m-\bar z)} ) }{ \big[( z-m)( m -\bar z)\big]^{ \frac{k}{2} +\frac{1}{4}} } 
  \bigg[-\frac{m}{2} \sqrt{\frac{z}{\bar z}} - \frac{m}{2} \sqrt{\frac{\bar z}{ z}}
+\sqrt{z \bar z}\bigg] .  \\  \nonumber
&\equiv& I_1 + I_2 +I_3.
\end{eqnarray}
Just as before the last line defines the 3 integrals required to obtain $\hat a(\Delta , l)$. 
Consider the $m$-th term of the first integral, 
\begin{eqnarray}
I_1(m)   &=& \pi m (-1)^{m}  K_l  \bigg(\frac{m_{\rm th}}{2\pi} \bigg)^{\frac{d}{2}} 
 \int_0^1 \frac{d\bar z}{\bar z} \int_1^{{\rm max } (m, \frac{1}{\bar z} )} \frac{dz}{z} \left[ ( z- \bar z)^{2\nu} ( z\bar z)^{\frac{1}{2} - \frac{\Delta}{2}} F_l \Big( \sqrt{ \frac{\bar z}{z} } \Big)  \sqrt{\frac{z}{\bar z}}   \right.  \nonumber \\
 && \qquad\qquad\qquad \qquad\qquad\qquad \left. 
 \times   \frac{J_{-\frac{1}{2} -k} ( m_{\rm th}  \sqrt{(z-m)(m-\bar z)} ) }{ \big[( z-m)( m -\bar z)\big]^{ \frac{k}{2} +\frac{1}{4}} }
 \right].
\end{eqnarray}
We  follow the same procedure as in the application of the inversion formula for the MFT to find the leading
poles. 
We first change variables as in (\ref{changver}), this leads to 
\begin{eqnarray}
I_1(m)   &=& \pi m^{k+ \frac{1}{2} -\Delta} (-1)^{m}  \bigg(\frac{m_{\rm th}}{2\pi} \bigg)^{\frac{d}{2}}  K_l
 \int_0^1 d\bar z  \int_1^{{\rm max } (\frac{1}{m\sqrt{\bar z}} , 1  )}  dz\; \left[ \bar z^{-1 -\frac{\Delta}{2} }  z^{ 2k -1 -\Delta} ( z-1)^{ -\frac{k}{2} - \frac{1}{4} } \right. \nonumber \\  &&  \qquad\;\; \left. \times ( 1-  \bar z)^{ 2k -1   } (1-z\bar z)^{-\frac{k}{2} - \frac{1}{4}}
  F_l( \sqrt{\bar z}) 
J_{-\frac{1}{2} -k} \big( m_{\rm th}  m  \sqrt{(z-1)(1- z \bar z)} \big)  \right].
\end{eqnarray}
Now we can expand in small $\bar z$, this decouples the integrals and then we perform the integral term by term in $\bar z$.
The leading  pole is given by 
\begin{eqnarray}
I_1(m)^{(0)} &=& 4\pi m^{k+\frac{1}{2} -\Delta}  (-1)^{m}    \bigg(\frac{m_{\rm th}}{2\pi} \bigg)^{ k +\frac{1}{2} }  K_l \frac{1}{ -\Delta + l + 2k - 1}
\\ \nonumber
&& \times \int_0^\infty y^{ - k + \frac{1}{2}} ( 1+ y^2)^{2k -1-\Delta}  J_{-\frac{1}{2} -k} \big( m_{\rm th} m \; y\big). 
\end{eqnarray}
To obtain the above equation, we have also made a change of variables to $y = \sqrt{ z-1}$.  The superscript denotes the fact 
that we are focussing on the leading term in the small $\bar z$ expansion. 
The integral over $y$ is known and we obtain 
\begin{eqnarray}
I_1^{(0)} &=&\sum_{m=1}^\infty
 \frac{ 2 (-1)^{m}\pi^{\frac{1}{2} -k} 2^{ -\Delta + k + \frac{1}{2} } m^{\frac{1}{2} -k} m_{\rm th}^{\Delta- k +\frac{1}{2}} K_l }{
\Gamma( 1+ \Delta - 2k) ( -\Delta + l + 2k -1) }  \;K_{\Delta - k + \frac{1}{2}} ( m_{\rm th} m ) .
\end{eqnarray}
We also need the first sub-leading term in the small $\bar z$ expansion.
After performing the $\bar z$ integral,  the first sub-leading contribution is given by 
\begin{eqnarray}
&&I_1^{(1)}(m)  = 4\pi m^{k +\frac{1}{2} - \Delta}(-1)^m    \bigg(\frac{m_{\rm th}}{2\pi} \bigg)^{ k +\frac{1}{2} }  K_l \frac{1}{ -\Delta + l + 2k + 1} \times {\cal I}, \\ \nonumber
&& {\cal I} = \int_0^\infty dy y^{-k +\frac{1}{2} } ( 1+y^2)^{2k -1-\Delta}  \left[
 \Big( -\frac{ (2k -1)(2+l) }{ 1+ 2k +2l }   + ( k +\frac{1}{2} ) ( 1+y^2) \Big) J_{-\frac{1}{2} -k} (y)  \right.  \\ \nonumber
&& \qquad \qquad \qquad\qquad\qquad\qquad\qquad\qquad \left.  + \frac{m m_{\rm th}}{2}  y ( 1+y^2) J_{\frac{1}{2} -k} (y)  \right].
\end{eqnarray}
Integrating over $z$ we obtain 
\begin{eqnarray}
&&I_1^{(1)}(m)  =
\frac{ 2 (-1)^{m}\pi^{\frac{1}{2} -k} 2^{ -\Delta + k + \frac{1}{2} } m^{\frac{1}{2} -k} m_{\rm th}^{\Delta- k +\frac{1}{2}} K_l }{
 ( -\Delta + l + 2k +1) } 
 \left[
  -\frac{ (2k -1)(2+l) }{ 1+ 2k +2l } \frac{K_{\frac{1}{2} +\Delta - k } ( m m_{\rm th}) }{\Gamma( 1+\Delta - 2k )} 
  \right. \nonumber \\
  &&\qquad\qquad\qquad
   \left.  + ( 2k + 1 )(m m_{\rm th})^{-1}  \frac{ K_{-\frac{1}{2} +\Delta - k } ( m m_{\rm th}) } {\Gamma( \Delta - 2k )}
   + \frac{ K_{-\frac{3}{2} +\Delta - k } ( m m_{\rm th}) } {\Gamma( \Delta - 2k )}
   \right].
\end{eqnarray}

Let us evaluate the  residue at the pole $\Delta = 2k + l   -1 $,  using the OPE expansion in (\ref{g2ope}) we can 
identify the one-point functions for the following operators 
\begin{eqnarray} \label{discgn}
\left. a_{{\cal O}_+}[n=0, l] \right|_{\rm disc} = \sum_{m=1}^\infty
 \frac{ 2 (-1)^{m}m^{\frac{1}{2} -k} 
 \pi^{\frac{1}{2}-k }K_l  m_{\rm th}^{l + k -\frac{1}{2}} }{ 2^{l + k -\frac{3}{2}} \Gamma(l)} K_{l + k-\frac{1}{2} } ( m m_{\rm th}).
\end{eqnarray}
To perform the sum over $m$, we use the following property of  the Bessel function with half integer orders. 
\begin{equation} \label{besli}
K_{l +\frac{1}{2}}(x) =  e^{-x} \sum_{n=0}^l \frac{ \sqrt{\pi} ( l + 1- n)_{2n}}{ (2x)^{n + \frac{1}{2}} n!}, \qquad l \in \mathbb{Z}.
\end{equation}
Substituting this identity in (\ref{discgn}) and performing the sum over $m$, we obtain 
\begin{equation} \label{gnonep}
a_{{\cal O}_+}[n=0, l] = \frac{l  }{ 2 \pi ^k (  k - \frac{1}{2} )_l} \left( \frac{m_{\rm th}}{2} \right)^{l +k -1} 
\sum_{n=0}^{l +k -1} \frac{ (l +k -n)_{2n}}{ (2 m_{\rm th}) ^n n!} {\rm Li}_{k+n} ( - e^{-m_{\rm th}} ) ,
\end{equation}
where $l = 2, 4, , \cdots$.  We will show subsequently that the contribution from the arcs at infinity vanishes for $l > 0$, 
Therefore we have identified  these residues to be the complete contribution to the one-point function of 
the operators ${\cal O}_+[0, l ]$. 
 A simple check is to observe that the one-point functions in (\ref{gnonep})  coincides with the MFT expression 
 in (\ref{opexpmft}) on taking $m_{\rm th}\rightarrow 0$ and taking $\Delta_\psi = k $ together with using the 
 relation (\ref{mftgnrel}). 

Similarly we can evaluate the contribution of the leading expansion in $\bar z$  in the integrands of 
 $I_2$ and $I_3$. This results in  poles at $\Delta = 2k + l +1$, the contributions are 
 \begin{eqnarray}
 I_2(m)  =
 \frac{ 2 (-1)^{m}\pi^{\frac{1}{2} -k} 2^{ -\Delta + k + \frac{1}{2} } m^{\frac{1}{2} -k} m_{\rm th}^{\Delta- k +\frac{1}{2}} K_l }{
 ( -\Delta + l + 2k +1)  \Gamma( 1+\Delta - 2k )} 
K_{\frac{1}{2} +\Delta - k } ( m m_{\rm th}),
  \end{eqnarray}
  and 
  \begin{eqnarray}
   I_3(m)  =
 \frac{ 8 (-1)^{m+1}\pi^{\frac{1}{2} -k} 2^{ -\Delta + k + \frac{1}{2} } m^{-\frac{1}{2} -k} m_{\rm th}^{\Delta- k -\frac{1}{2}} K_l }{
 ( -\Delta + l + 2k +1)  \Gamma( \Delta - 2k )} 
K_{-\frac{1}{2} +\Delta - k } ( m m_{\rm th}) .
  \end{eqnarray}
  The 
 sum of the residues of the poles from $I_1^{(1)}(m), I_2(m)$ and $I_3(m)$ is given by 
  \begin{eqnarray}
  && -I(m)|_{{\hbox{ Res at}}\; \Delta = 2k + l +1} =  \frac{(-1)^m K_l \pi^{\frac{1}{2} - k} 2 ^{ - k - l +\frac{1}{2} }
  m^{-\frac{1}{2} - k } (m_{\rm th} )^{l + k + \frac{1}{2} } 
}{\Gamma( l +1)}   \\ \nonumber
  && \times \left[  \frac{3 - 2k}{ 1+ 2k + 2l } ( m m_{\rm th}) K_{l +k + \frac{3}{2} } ( m m_{\rm th} ) 
  + ( 2k -3) K_{l +k + \frac{1}{2}} ( m m_{\rm th} ) 
  + ( m m _{\rm th}) K_{l + k -\frac{1}{2} } ( m m_{\rm th}) 
  \right].
  \end{eqnarray}
  Then using the recurrence relation of the Bessel functions,
 \begin{equation}
 \frac{2\nu}{x} K_{\nu} (x) = -K_{\nu -1} + K_{\nu +1},
 \end{equation}
   we can simply this to 
  \begin{eqnarray}
   - I(m)|_{{\hbox{ Res at}} \;\Delta = 2k + l +1} &=&  \frac{ (-1)^{m} K_l \pi^{\frac{1}{2} - k }
    2^{-k-l+\frac{3}{2}} m^{\frac{1}{2} -k} {m_{\rm th}}^{l +k +\frac{3}{2}}  
    (l+2) }{\Gamma( l +1)  ( 1 + 2k + 2l ) } \nonumber \\ 
   && \qquad\qquad\qquad \times  K_{l + k - \frac{1}{2}} ( m m_{\rm th})  .
  \end{eqnarray}
  We can sum over $m$ using the identity (\ref{besli}) which results in the complete contribution  from the discontinuity across 
  the cuts to the residues at $\Delta = 2k + l +1$. 
  \begin{eqnarray} \label{gnadiscg}
 - \hat a( \Delta, l )_{\rm disc}|_{{\hbox{ Res at}} \;\Delta = 2k + l +1}  &=&   \frac{ ( l+2) }{  \pi^k ( 1+ 2k + 2l ) ( k - \frac{1}{2} )_l } 
 \left( \frac{m_{\rm th}}{2} \right)^{l +k + 1 }  \\ \nonumber 
& & \quad\quad\times   \sum_{n=0}^{l +k -1} \frac{ (l +k -n)_{2n}}{ (2 m_{\rm th}) ^n n!} {\rm Li}_{k+n} ( - e^{-m_{\rm th}} ).
  \end{eqnarray} 
  Finally  for $l>0$, from the residues at $\Delta = 2k +l +1$ given in (\ref{gnadiscg}) and the OPE expansion in (\ref{g2ope}), we 
  obtain 
  \begin{eqnarray} \label{equationgnop}
  	&& a_{{\cal O}_+}[1, l] + \frac{ 2( l +1)}{ ( l +2) ( 2k + 1 + l ) } a_{{\cal O}_-}[0, l] 
  	=    \\ \nonumber
  	&&\qquad\qquad \frac{ ( l+2) }{  \pi^k ( 1+ 2k + 2l ) ( k - \frac{1}{2} )_l } 
  	\left( \frac{m_{\rm th}}{2} \right)^{l +k + 1 } 
  	\sum_{n=0}^{l +k -1} \frac{ (l +k -n)_{2n}}{ (2 m_{\rm th}) ^n n!} {\rm Li}_{k+n} ( - e^{-m_{\rm th}} ),
  \end{eqnarray}
  where again $l=0,2,\cdots$.
  \subsubsection*{Contribution from the arcs }
  
  To complete the evaluation of the one-point function we need to evaluate the contribution from the arc at 
  infinity. 
  This is given by  the expression in (\ref{contriarc}).  Taking the limit $w\rightarrow\infty$ in $g_2(x)$, it can be seen only the 
  mode $m=0$ contributes, this is due to fact the argument of the Bessel function has a square root, which ensures that 
  it vanishes on the arc exponentially for $m>0$.
  Furthermore due to the integration over the full circle in the $w$ plane only $l=0$ contributes. 
  Due to these reasons the contribution from the arcs is given by 
  \begin{eqnarray}
  \hat a_{\rm arc} (\Delta, 0)  = - 4\pi K_0\left( \frac{m_{\rm th} }{2\pi} \right)^{\frac{d}{2}}
   \int_0^1 dr  \frac{K_{ k + \frac{1}{2} } (m_{\rm th} r) } {r^{\Delta - k +\frac{1}{2} } }  .
  \end{eqnarray}
  To obtain the location of the pole in the $\Delta$ plane we can push the  upper limit of the integral to $\infty$. This does not change either the location of the poles nor their residues.
  This results in the following 
  \begin{equation}
    \hat a_{\rm arc} (\Delta, 0)  =- \frac{1}{4\pi^{k +\frac{1}{2}}} \big(  \frac{m_{\rm th}}{2}\big)^\Delta 
    \Gamma\Big( -\frac{\Delta}{2} \Big) 
    \Gamma\Big( \frac{1}{2} - \frac{\Delta}{2} + k \Big) .
  \end{equation}
  We therefore obtain the residue at $\Delta = 2k +1$
  \begin{equation} \label{gnarcg}
  \hat a(\Delta, 0)_{\rm arc}|_{{\hbox{ Res at}} \;\Delta = 2k + 1}
   = - \frac{1}{2\pi^{k + \frac{1}{2} } } \Big( \frac{m_{\rm th}}{2}  \Big)^{ 2k + 1} \Gamma\Big( - k - \frac{1}{2}  \Big) .
  \end{equation}
  
  \subsubsection*{Gap equation from OPE inversion}

Similar to the analysis \cite{Iliesiu:2018fao} for the bosonic  critical $O(N)$ model,  let us demand that the low lying operator 
${\cal O}_{-}[l=0, n=0]$  with $\Delta = 2k +1$ does not occur in the spectrum. 
These are the operators which are schematically of the form $\bar\psi\gamma^\mu \partial_\mu \psi$. 
This implies that contribution to the residues  from the discontinuity of the branch cuts in (\ref{gnadiscg}) together with the 
arcs at infinity in (\ref{gnarcg}) at the pole $\Delta = 2k+1$ must vanish. 
\begin{eqnarray} \label{gngapa}
\left[ - \hat a( \Delta, 0 )_{\rm disc}-\hat a(\Delta, 0)_{\rm arc} \right]|_{{\hbox{ Res at}} \;\Delta = 2k + 1}  =0.
\end{eqnarray}
Note that this combination is the one-point function   $a_{{\cal O_-}}[0,0]$ as can be seen from the 
OPE  expansion (\ref{g2ope}). 
Using (\ref{gnadiscg}) and (\ref{gnarcg}) in (\ref{gngap}) we obtain the equation
\begin{eqnarray} \label{gngap1}
2 ( 2m_{\rm th})^k \sum_{n=0}^{k-1} \frac{ ( k-n)_{2n} }{ ( 2m_{\rm th})^n n! }  {\rm Li}_{k+n} ( - e^{-m_{\rm th} })
+ \frac{(m_{\rm th}) ^{2k}  \sqrt{\pi}}{\Gamma( \frac{1}{2} + k) \cos\pi k} =0,
\end{eqnarray}
which precisely agrees with that obtained from the partition function in (\ref{gngap}). 

It is important to mention that this gap equation was also obtained \cite{Petkou:2018ynm}
 for the Gross-Neveu model  
by demanding operators ${\cal O}_0[0,0]$ which are 
schematically of the form $\bar\psi \psi$  \footnote{See equation (27) of \cite{Petkou:2018ynm}.}.
We will review this subsequently.    Heuristically this fact could have been anticipated, at the 
large $N$ saddle point $\bar\psi \gamma^\mu \partial_\mu \psi \sim  \zeta \bar\psi\psi = m_{\rm th} \bar \psi\psi$ 
where $\zeta$ is the field introduced by the Hubbard-Stratonovich transformation to linearise the 4-fermi interaction. 
Therefore the
vanishing of the one-point function  ${\cal O}_0[0,0]$  implies that the one-point function 
${\cal O}_{-}[0,0]$ also vanishes. 
The fact that the explicit computation does indeed bear out this expectation 
 is an important consistency check of the OPE inversion formula developed for the correlator $g_2(x)$.

  \subsection{OPE inversion on  $g_3(x)$}
  
  It can be easily seen that  using the definition of $g_3(x)$ in (\ref{corr}) and (\ref{mth2pt}) , this correlator  is given by 
  \begin{align}
	g_3(x)=\sum_{n, k_0 = 
	2\pi ( n + \frac{1}{2}) } \partial^2 \left[ 
	\int \frac{d^{d-1}x}{(2\pi)^{d-1}}  \bigg(\frac{1}{k^2+m_{th}^2}\bigg) e^{ikx}
	\right].
\end{align}
We can use the equation (\ref{poiresum}) which relates the term in the square brackets to the Bessel function to take the 
derivatives. This leads to 
\begin{align}
	g_3(x)=\sum_{m\in \mathbb{Z}}\frac{(-1)^m}{(2 \pi )^{\frac{d}{2}}} m_{th}^{\frac{d}{2}+1} |x^{(m)}|^{\frac{2-d}{2}} K_{\frac{d-2}{2}}\left(m_{th} |x^{(m)}|\right) .
\end{align}
Thus $g_3(x)$ is a simpler correlator and similar to that encountered in the bosonic $O(N)$ model. 
We can apply the inversion formula  as before
\begin{eqnarray} \label{gninverfg3}
\hat a_{\rm disc} ( \Delta, l ) = 2 K_l \int_0^1 \frac{d\bar z}{\bar z} \int_1^{\frac{1}{\bar z} } \frac{dz}{ z}
( z- \bar z)^{2\nu} ( z\bar z)^{\frac{1}{2} - \frac{\Delta}{2}} F_l \Big( \sqrt{ \frac{\bar z}{z} } \Big) {\rm Disc}[g_3(z, \bar z) ] ,
\end{eqnarray}
where $l$ is even.
%
Since $g_3(x)$ is still a sum of BesselK functions, we can  use (\ref{bcrelation}) to obtain the discontinuity across the 
branch cuts. Proceeding along the similar lines, the contribution to the residue  of  $\hat a_{\rm disc} ( \Delta, l ) $ at 
the poles $\Delta = 2k + l $ is given by  
\begin{eqnarray}
-\hat a_{\rm disc} ( \Delta, l )\big|_{{\mbox{ Res at} }\; \Delta = 2k +l } = \frac{2}{\pi^k ( k + \frac{1}{2} )_l} 
\big( \frac{m_{\rm th}}{2} \big)^{l +k +1}  \sum_{n=0}^{k +l -1}  \frac{ (l +k -n)_{2n}}{ (2 m_{\rm th}) ^n n!} {\rm Li}_{k+n} ( - e^{-m_{\rm th}} ). \nonumber\\
\end{eqnarray}
Similar to the discussion for $g_2(x)$, the residue at $\Delta =2k, l =0$ receives contribution from the arcs at infinity
which is given by
\begin{align}
-\hat a_{\rm arc} ( \Delta, 0 )\big|_{{\mbox{ Res at} }\; \Delta = 2k  } =
\big( \frac{m_{\rm th} }{2} \big) ^{2k +1} \frac{\Gamma \left(- k + \frac{1}{2} \right)}{\pi^{k + \frac{1}{2} }}.
\end{align}
Now the expectation value $a_{{\cal O}_-}[0,0]$ which refers to the operator schematically of the 
form $\bar\psi \gamma^\mu \partial_\mu \psi$  is given by the combination 
\begin{eqnarray}\label{gapominus}
&&a_{{\cal O}_-}[0,0] = -\big[ \hat a_{\rm disc} ( \Delta, 0 ) +a_{\rm arc} ( \Delta, 0 )\big]\Big|_{{\mbox{ Res at} }\; \Delta = 2k  },
\\ \nonumber
&=&   \frac{2}{\pi^k} 
\big( \frac{m_{\rm th}}{2} \big)^{ k +1}  \sum_{n=0}^{k  -1}  \frac{ (k -n)_{2n}}{ (2 m_{\rm th}) ^n n!} {\rm Li}_{k+n} ( - e^{-m_{\rm th}} ) + \big( \frac{m_{\rm th} }{2} \big) ^{2k +1} \frac{\Gamma \left(- k + \frac{1}{2} \right)}{\pi^{k + \frac{1}{2} }} ,
\\ \nonumber
&=& 0.
\end{eqnarray}
Again, demanding that this operator does not exist in the spectrum results in the same gap equation  (\ref{gngap1})
obtained by considering the correlator $g_2(x)$. 
This agreement is necessary for the consistency of obtaining the expectation value $a_{{\cal O}_-}[0,0]$ both from 
$g_2(x)$ and $g_3(x)$. 
Finally using the OPE expansion of $g_3(x)$  in (\ref{g3 ope}), (\ref{g3}),  we see that 
poles at  $\Delta = 2k$ correspond to the coefficient $c[0, 0]$. Then from the relations in (\ref{cg3}), we obtain the 
linear relation between the expectation values $a_{{\cal O}_-}[0,0]$ and $a_{{\cal O}_+}[0,0]$
\begin{align}\label{eq2GN}
	2a_{\mathcal{O}_+}[1,l]+(l+1)a_{\mathcal{O}_-}[0,l]=
	\frac{2}{\pi^k ( k + \frac{1}{2} )_l} 
\big( \frac{m_{\rm th}}{2} \big)^{l +k +1}  \sum_{n=0}^{k +l -1}  
\frac{ (l +k -n)_{2n}}{ (2 m_{\rm th}) ^n n!} {\rm Li}_{k+n} ( - e^{-m_{\rm th}} ) .
\end{align}
We can use the equations (\ref{equationgnop}) and (\ref{eq2GN}), to obtain values $a_{\mathcal{O}_+}[1,l]+$ and $a_{\mathcal{O}_-}[0,l]$, but it is not 
illustrative. 
What is important to note  is that on comparing the LHS of (\ref{equationgnop}) and (\ref{eq2GN}) and the expectation value 
 (\ref{gnonep}), we see that  for a given $l$, $a_{\mathcal{O}_+}[1,l]+$ and $a_{\mathcal{O}_-}[0,l]$ are proportional to 
 $m_{\rm th}^2 a_{\mathcal{O}_+}[0,l]$.

  \subsection{OPE inversion on  $g_1(x)$}
  
  Finally for completeness, let us give the results for $g_1(x)$, the correlator studied in \cite{Petkou:2018ynm}. 
  From  (\ref{corr}) and (\ref{gnftcor1}), we see that
  \begin{align}
	g_1(x)=\sum_{n, k_0 = 2\pi ( n + \frac{1}{2} ) }
	\int \frac{d^{d-1}x}{(2\pi)^{d-1}}  \bigg(\frac{m_{th}}{k^2+m_{th}^2}\bigg) e^{ikx}.
\end{align} 
Performing the Poisson re-summation and then evaluating the Fourier transform we obtain 
\begin{align}
	g_1(x)=\sum_{n\in \mathbb{Z}}\frac{(-1)^n}{(2 \pi )^{\frac{d}{2}}} m_{th}^{\frac{d}{2}} |x^{(n)}|^{\frac{2-d}{2}} K_{\frac{d-2}{2}}\left(m_{th} |x^{(n)}|\right).
\end{align}
From the OPE expansion in (\ref{opexp1}), we see that this correlator contains the information of the 
one-point functions of the operators ${\cal O}_0[0, l]$. These operators are schematically represented by the 
traceless symmetric 
bi-linears given in (\ref{fermbilinear}).  
Again using the OPE inversion formula on $g_1(x)$, the residues at poles $\Delta = 2k +l$  lead to the following 
one-point function
\begin{align}\label{ozeroexp}
	a_{\mathcal{O}_0}[0,l]&= \frac{ 1}{ \pi^k (k+\frac{1}{2})_l } \big( \frac{m_{\rm th}}{2} \big)^{l +k}
	\sum_{n=0}^{l+k-1}\frac{  (l+k-n)_{2 n} \text{Li}_{k+n}\left(-e^{-m_{th}}\right)}{ (2m_{\rm th})^{n} }, \\ \nonumber
	& l = 2, 4, \cdots.
\end{align}
For $l = 0$, as before there is a contribution to the residue both from the arc as well as the disc. 
The residue at the $\Delta =2k$ pole is given by
\begin{eqnarray}
-\hat a_{\rm arc} ( \Delta, 0 )\big|_{{\mbox{ Res at} }\; \Delta = 2k  } = \frac{ m^{2k}}{2^{2k+1} \pi^{k +\frac{1}{2}}} 
\Gamma( \frac{1}{2} - k ).
\end{eqnarray}
Combining the contribution of the arc and the disc results in the expectation value of the operator $\bar\psi\psi$. 
Demanding that this operator does not exist in the spectrum results in the gap equation 
\begin{eqnarray}\label{gap0}
&&a_{{\cal O}_0} [0,0] = -\big[ \hat a_{\rm disc} ( \Delta, 0 ) +\hat a_{\rm arc} ( \Delta, 0 )\big]\Big|_{{\mbox{ Res at} }\; \Delta = 2k  }=0,
\\ \nonumber
&&{\rm thus,}\qquad \frac{ 1}{ \pi^k } \big( \frac{m_{\rm th}}{2} \big)^{k}
	\sum_{n=0}^{k-1}\frac{  (k-n)_{2 n} \text{Li}_{k+n}\left(-e^{-m_{th}}\right)}{ (2m_{\rm th})^{n} } 
	+ \frac{ m^{2k}}{2^{2k+1} \pi^{k +\frac{1}{2}}} 
\Gamma( \frac{1}{2} - k )
=0.
\end{eqnarray}

\subsubsection*{Properties of the one-point functions in the GN model}

Comparing the equation obtained by demanding the operators ${\cal O}_-[0, 0]$ and ${\cal O}_0[0,0]$ do not exist, equations 
 (\ref{gapominus}) and \eqref{gap0} respectively, we see that the equations are related by just an overall multiplicative factor of $m_{\rm th}$. 
 Therefore, the gap equation is same and one obtains no new conditions which is important for the consistency of starting 
 with the thermal propagator with one parameter $m_{\rm th}$. 
 As explained earlier, this might have been expected by the large $N$ saddle point equations of motion $\bar \psi \gamma^\mu\partial_\mu \psi  \sim m_{\rm th} \bar\psi \psi$. 
 What is perhaps  more non-trivial is the following observation. 
Comparing  the one-point functions of operators ${\cal O}_+[0, l]$  in (\ref{gnonep}) and  operators ${\cal O}_0[0, l]$  in
(\ref{ozeroexp}) 
we see that
\begin{equation} \label{nicrel}
a_{{\cal O}_0}[0, l] = m_{\rm th} a_{{\cal O}_+}[0,l], \qquad\qquad l = 2, 4, \cdots.
\end{equation}
Here there is no obvious equation of motion relating these expectation values.  Such a relation must be specific to the
critical Gross-Neveu model at large $N$. 
Observe that for the MFT of fermions the correlator $g_0(x)$ vanishes, so all expectation values 
$a_{{\cal O}_0}[0, l]$ vanish. However the  expectation values  $a_{{\cal O}_+}[0,l]$ are non-trivial and are given by 
(\ref{opexpmft}).  Of course for the MFT of free fermions, we have $m_{\rm th}=0$, and therefore the fact that 
$a_{{\cal O}_0}[0, l]$ vanishes and $a_{{\cal O}_+}[0, l]$ does not, is consistent with (\ref{nicrel}). 

For the  critical Gross-Neveu model it is likely there are  more relations of the kind (\ref{nicrel}). 
From the observations made after equation (\ref{eq2GN}), we know that the expectation values 
$a_{{\cal O}_+}[1,l]$ and $a_{{\cal O}_-}[0,l]$ are also proportional to $m_{\rm th}^2 a_{{\cal O}_+}[0,l]$.  Here the proportionality constants involve spin and numerical factors. 
 This and 
the relation (\ref{nicrel}) hint 
 that the expectation values of   bilinears of the form ${\cal O}_+[n, l ], n>0$, 
${\cal O}_-[n, l], n\geq 0$ and ${\cal O}_{0}[n, l], n\geq 0$ are all related to the expectation value  $ a_{{\cal O}_+}[0,l]$. 
It will be interesting to prove this.

\section{Large $d$ and large spin behaviour of one-point functions}\label{seclargds}

As we have mentioned in the introduction, since the OPE inversion formula provides  compact expressions for 
one-point functions, we study their behaviour at large dimensions $d$ and spin $l$. 
As we have discussed in section \ref{secinv}, the one-point function $a_{\cal O}$ is proportional  to the thermal expectation 
value of the corresponding operator.  The proportionality constants involve the structure constant, $f_{\bar\psi\psi {\cal O}}$ and 
the normalization of the two-point function of ${\cal O}$, $c_{\cal O}$. 
To eliminate this dependence, we  use the following, 
for the  $O(N)$ model or the Gross-Neveu models OPE coefficients, anomalous dimensions 
  are the same both in the Gaussian fixed point as well as the critical fixed point  at large $N$\cite{Vasiliev:1981dg,Lang:1992zw,Petkou:1994ad,Petkou:1995vu,Moshe:2003xn,Fei:2014yja}. 
  Therefore,  we study the 
ratio of the one-point functions $a_{\cal O}$ at these fixed points. 
\begin{align}
	r(l,k)= \frac{ a_{\cal O}[l]_{ m_{\rm th} \neq 0 , k }}{  a_{\cal O}[l]_{ m_{\rm th} =0, k  }}.
\end{align}
Here
$m_{\rm th}\neq 0, k$ denotes  evaluating the one-point function for 
the real positive solution of the gap equation at dimension $d=2k+1$.
We discuss two cases, first we examine the behaviour of $r(l,k)$ by increasing $k$ keeping $l$ fixed and in the second case we study $r(l,k)$ by increasing $l$ at fixed $k.$

In the rest of the section we study the one-point function of the stress tensor in detail for the two theories. 
The energy density for conformal field theories on $S_1\times R^{d-1}$ can be written as
\begin{eqnarray} \label{defcmd}
T_{00} = - E =   \frac{  (d-1) \Gamma( \frac{d}{2} )}{\pi^{\frac{d}{2}} \beta^d} \times   c(m_{\rm th}, d ) .
\end{eqnarray}
For free bosons $c (m_{\rm th} =0, d ) = -\zeta(d)$, which is the Stefan-Boltzmann value \footnote{We are examining the energy density divided by $N$.}. 
Therefore, we can consider $c(m_{\rm th}, d )$ as a rough measure of the degrees of freedom which is seen on heating the system. 
We plot the $c(m_{\rm th}, d)$ as a function of $d$, we see that   $c(m_{\rm th}, d)$ vanishes as $d$ increases. 
For  fermions  we can further write 
$c(m_{\rm th}, d ) = 2^\frac{d-1}{2}  \tilde c(m_{\rm th}, d ) $, where the factor $2^{\frac{d-1}{2}} $ is due to the
dimension of the spinor in odd $d$ dimensions. 
For free fermions, the Stefan-Boltzmann value is  $ \tilde c(m_{\rm th}, d ) = - ( 1- 2^{-( d-1)}) \zeta( d) $. 
Again we see that the $\tilde c( m_{\rm th}, d ) $ vanishes as $d$ increases,  that is  the degrees of freedom seen 
by heating the system again decreases.

\subsection{ $O(N)$ model}

\begin{figure}[t]
	
	\begin{subfigure}{.475\linewidth}
		\includegraphics[width=\linewidth]{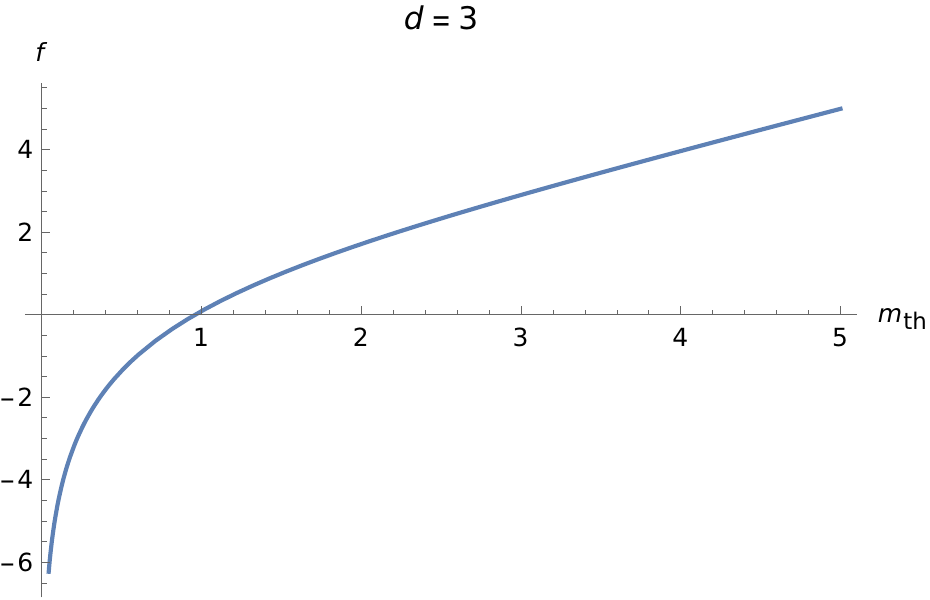}
		\caption{}
	\end{subfigure}\hfill 
	\begin{subfigure}{.475\linewidth}
		\includegraphics[width=\linewidth]{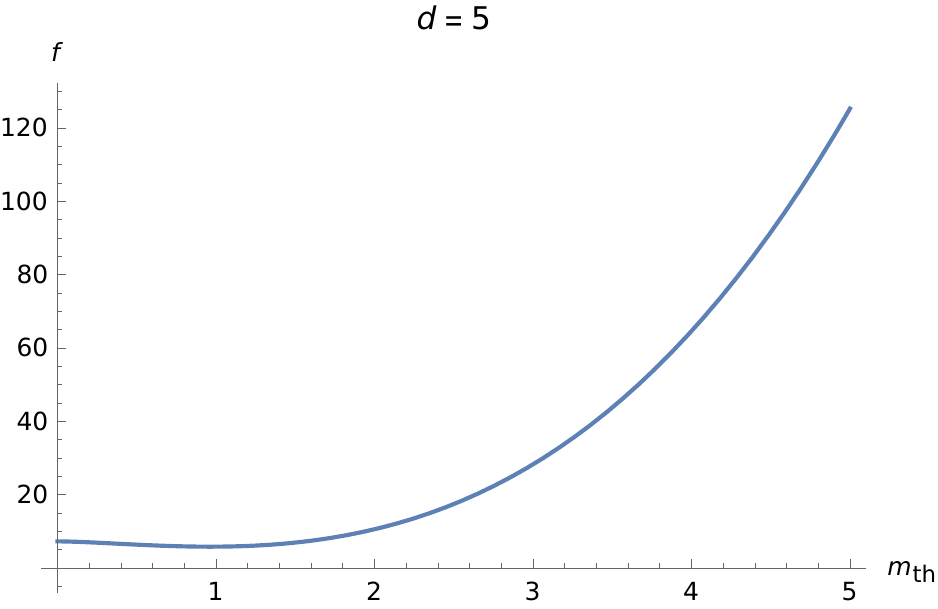}
		\caption{}
	\end{subfigure}
	\par\bigskip
	\par\bigskip
	\begin{subfigure}{.475\linewidth}
		\includegraphics[width=\linewidth]{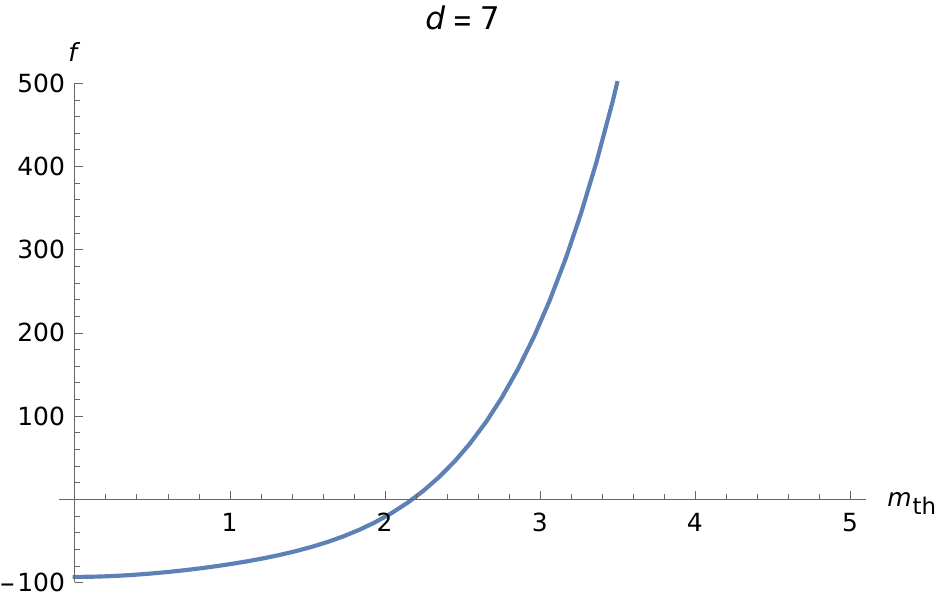}
		\caption{}
	\end{subfigure}\hfill 
	\begin{subfigure}{.475\linewidth}
		\includegraphics[width=\linewidth]{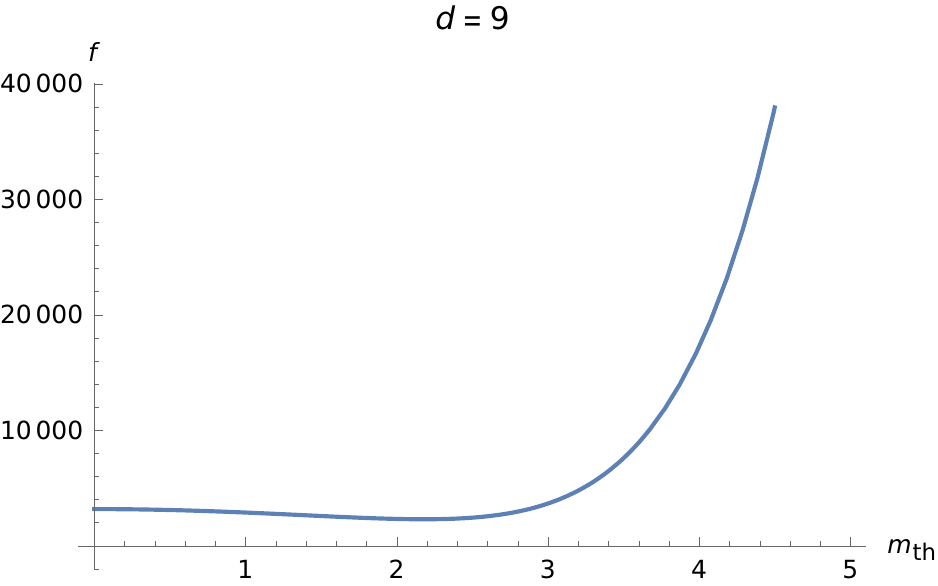}
		\caption{}
	\end{subfigure}
	
	\caption{The LHS of the gap equation \eqref{gapeqb} for the 
	$O(N)$ model (denoted by $f$ in the figure) has been plotted as a function of $m_{\rm th}$, for $d=3$ and $d=7$ the graph cuts the $x$-axis only once while for $d=5,9$, it is always positive. Similar trend follows for higher values of $ d $, which is possible to check numerically. }
	\label{fig:roc}
\end{figure}

In this section we will numerically study the behaviour of one-point functions for $O(N)$ model with increasing $ d $
 at a fixed  $l$ and vice versa. For this analysis, one should first look for the existence of real solutions of the gap equation at various odd dimensions.
The gap equation for the $O(N)$ model at strong coupling in odd dimension can be derived using the standard field theoretic technique elaborated in appendix  \ref{appendgap}. 
The same gap equation is obtained in \cite{Petkou:2018ynm} by demanding the absence of the operator $ \phi^2 $ in the ope of the two-point function for critical $O(N)$ model using the formalism of inversion formula.
\begin{align}\label{gapeqb}
	(m_{\rm th}) ^{2 k}+{2 \sqrt{\pi } \sum _{m=0}^{k-1} \frac{ (2m_{\rm th}) ^{k-m} (k-m)_{2m} \text{Li}_{k+m}\left(e^{-m_{\rm th} }\right)}{\Gamma \left(\frac{1}{2}-k\right)m!}}=0.
\end{align}
For $d=3,7,11,15,\ldots$, the gap equation is observed to have only one positive real solution for $m_{\rm th}$ and for $d=5,9,13,\ldots$ no real solution for $m_{\rm th}$ exists. We have shown this in figure \ref{fig:roc}. This phenomenon was 
noticed in  \cite{Petkou:2018ynm}. 

The one-point function for double twist operator of kind $ \phi\partial_{\mu_1}\cdots\partial_{\mu_l}\phi $ is given by \cite{Petkou:2018ynm},
\begin{align}
	a_l=\frac{(1+(-1)^l)}{2^{2 l+k}l!(k-\frac{1}{2})_l}\sum _{n=0}^{k+l-1} \frac{2^{n+1} m_{th} ^n (2 (k+l-1)-n)! \text{Li}_{2 k-n+l-1}\left(e^{-m_{th} }\right)}{n! (k-n+l-1)!}.
\end{align}
The ratio of one-point functions at  the non-trivial solution of the gap equation  to the same at $m_{\rm th}=0$, 
the Stefan-Boltzmann value is  given by
\begin{align}
	r_b(l,k)=\frac{a_l|_{m_{\rm th}\neq 0 , k }}{a_l|_{m_{\rm th}=0, k }}.
\end{align}
$m_{\rm th}\neq 0, k $ denotes the real positive solution of the gap equation at dimension $d=2k+1$
can be obtained  by solving the gap equation numerically. 

\begin{figure}[h]
	\begin{center}
		\includegraphics[scale=.7]{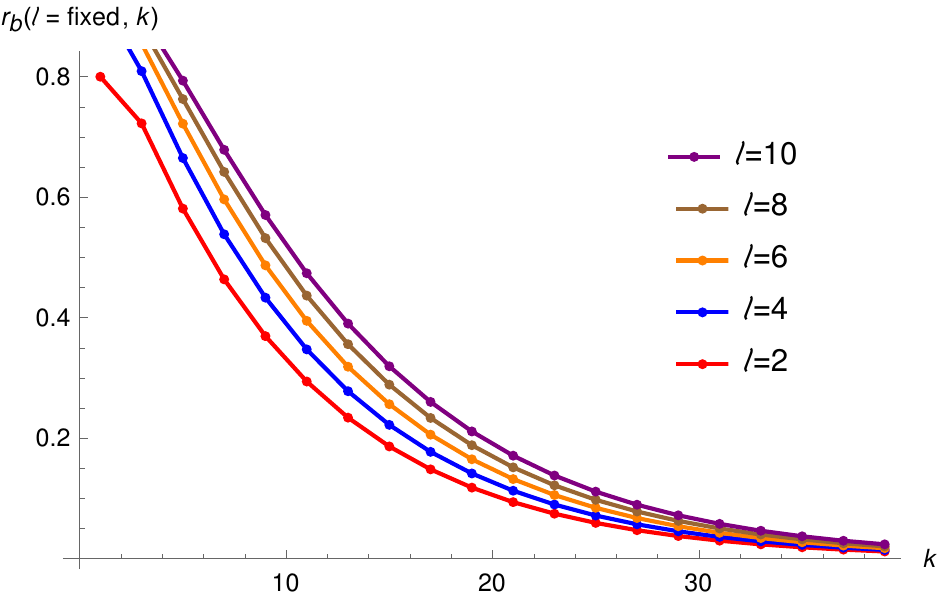}
	\end{center}
	\caption{The ratio $ r_b(l,k) $ for the $O(N)$ model
	 is plotted against $ k $($\in$ odd integers) for various fixed values of spin $ l $. The plot shows that the one-point functions at fixed $l$ becomes smaller in comparison with the same for the free theory with increasing $ d $.}
	\label{fig: ratio for bosons}
\end{figure}

\begin{figure}[h]
	\centering
	\includegraphics[scale=.7]{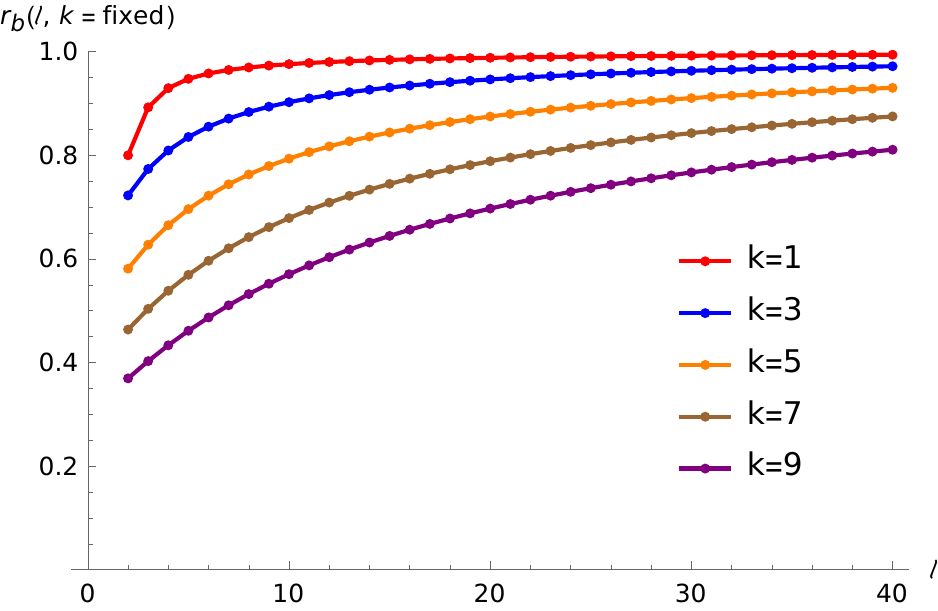}
	\caption{Plot showing the nature of the ratio $r_b(l,k)$ for the $O(N)$ model versus $l$, keeping $k$ fixed. } 
	\label{large J bosons}
\end{figure}

We examine the behaviour of $r_b(l,k)$ with increasing $k$ at fixed $l$; we restrict $ k $ to be odd, as the real solution for 
$m_{\rm th}$ exists only for odd $ k $. The result of this analysis is shown in figure \ref{fig: ratio for bosons}. The observation is that the ratio of the one-point function of double twist operator at fixed value of spin  $l$ evaluated at the Gaussian fixed point to that at the free theory limit keeps on decreasing with increasing $k$, but this ratio falls slower for higher values of $l$

\begin{align}
	\lim_{k\to \infty} r_b(l={\rm fixed},k)\to 0.
\end{align}
Now, we keep the $ k $ fixed and increase $ l $ and observe how $ r_b(l,k) $ behaves. The result of our numerical study is described in figure \ref{large J bosons}. At a fixed $k$, $r_b(l,k)$ saturates at 1 for large values of $l$, which indicates that the 
one-point function of large spin operator evaluated at the critical point of $O(N)$ model is equal to that in free theory.
\begin{align}
	\lim_{l\to\infty}r_b(l,k={\rm fixed}) =1.
\end{align}

\begin{figure}[th]
	\centering
	\includegraphics[scale=.7]{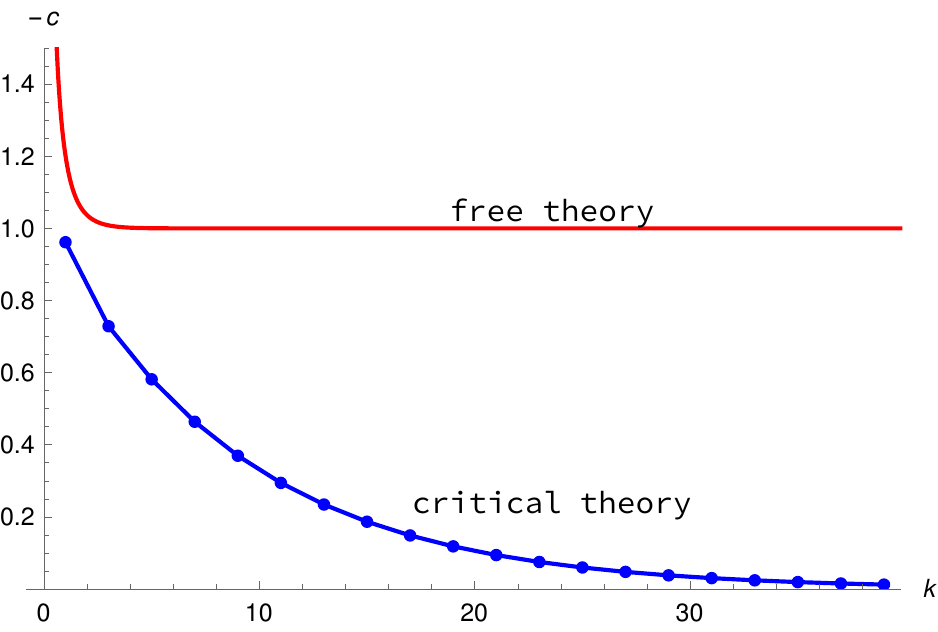}
	\caption{Plot of $c(m_{\rm th}, k )$ for the free theory and the critical theory in the $O(N)$ model.}
	\label{f bosons}
\end{figure}

Let us  study the behaviour of the coefficent $c( m_{\rm th}, k )$ which determines 
energy density as defined in (\ref{defcmd}). 
This coefficient can be extracted from the following expression for the energy density which can be derived from the 
partition function.
\begin{align}
	 E=\sum _{m=0}^{k+1} \frac{m_{\rm th} ^{k-m+1} \left((k+m)^2+k-m\right) \Gamma (k+m) \text{Li}_{k+m}\left(e^{-m_{\rm th}   }\right)}{\pi^k 2^{k+m-1}m! \Gamma (k-m+2)}-\frac{m_{\rm th} ^{2 k+1} \Gamma \left(-k-\frac{1}{2}\right)}{\pi^{k+\frac{1}{2}}2^{2k+1}}.
\end{align}
The energy density at the Gaussian fixed point is obtained by taking $m_{\rm th}\to 0$ in the above expression, this results in 
\begin{align}
-c( m_{\rm th}=0, k ) = \zeta( 2k +1) .
\end{align}
At the non-trivial fixed point this coefficient can be evaluated numerically. 
In figure (\ref{f bosons}), we plot this measure of degrees of freedom for the critical fixed point 
and also the Stefan-Boltzmann value for reference.  Note that the $-c( m_{\rm th}=0, k ) $ tends to zero 
for the critical point while the Stefan-Boltzmann value tends to one for large dimension $d$. 
\subsection{Gross-Neveu model}
\begin{figure}[hbt!]
	
	\begin{subfigure}{.475\linewidth}
		\includegraphics[width=\linewidth]{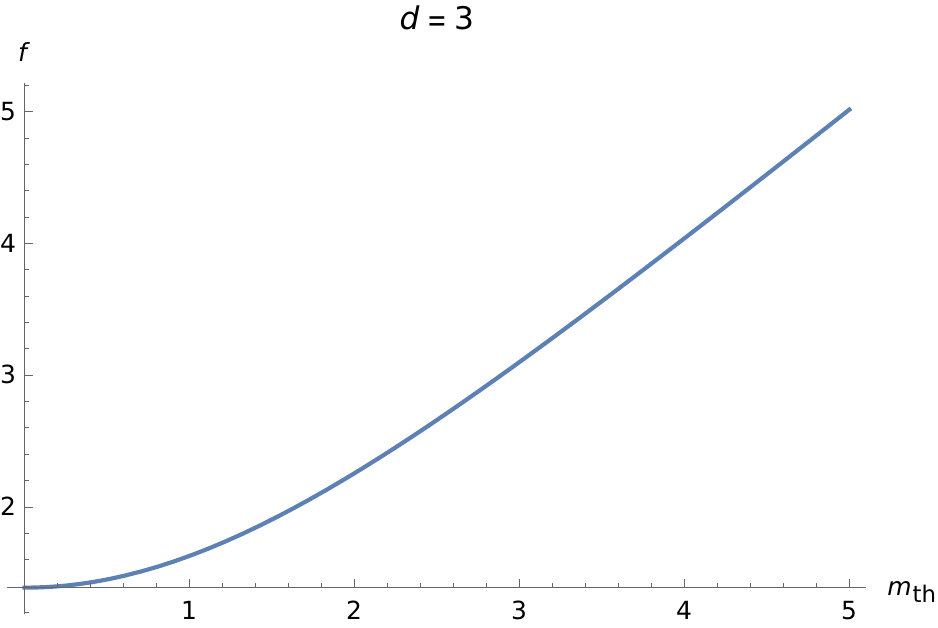}
		\caption{}
		\label{MLEDdet}
	\end{subfigure}\hfill 
	\begin{subfigure}{.475\linewidth}
		\includegraphics[width=\linewidth]{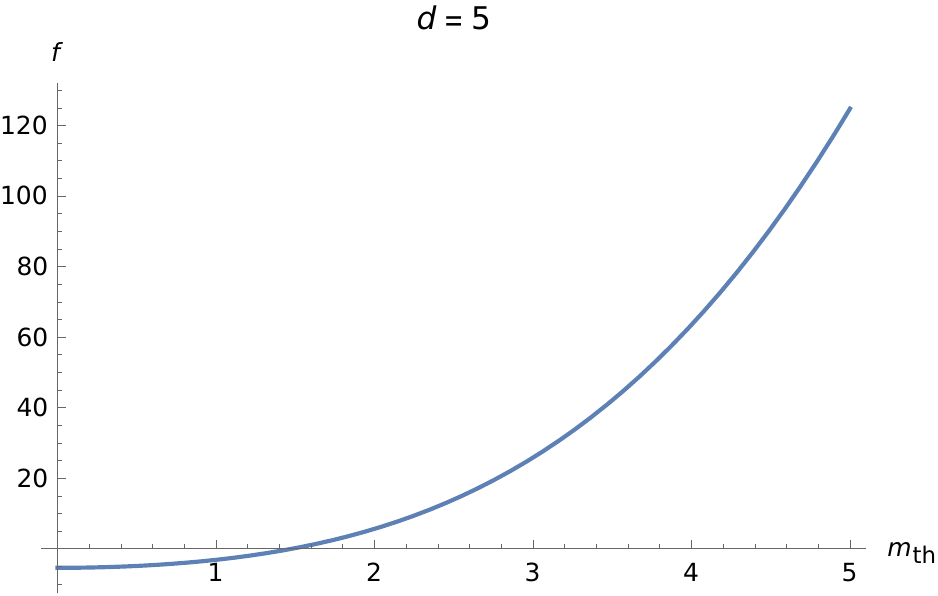}
		\caption{}
		\label{energydetPSK}
	\end{subfigure}
	\par\bigskip
	\par\bigskip
	\begin{subfigure}{.475\linewidth}
		\includegraphics[width=\linewidth]{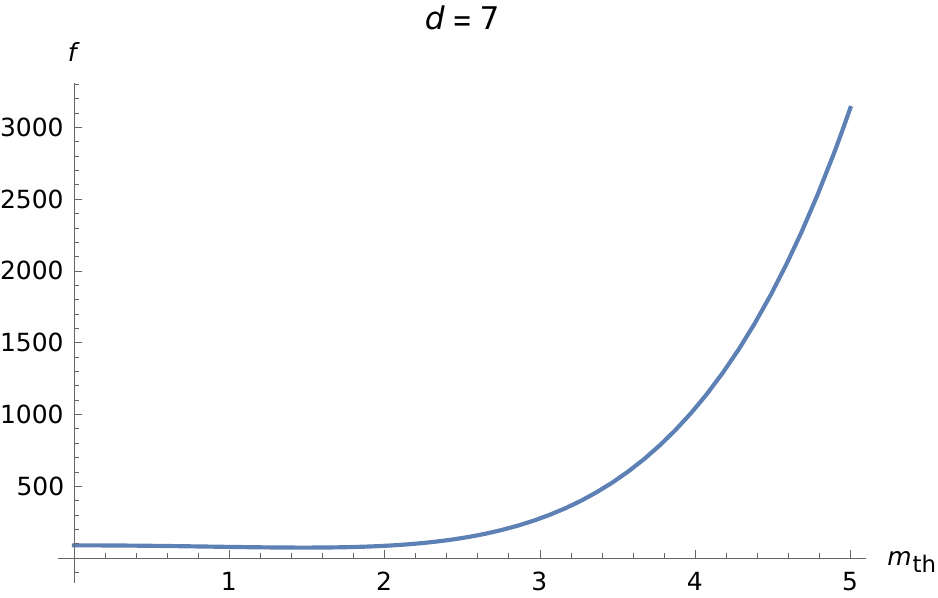}
		\caption{}
		\label{velcomp}
	\end{subfigure}\hfill 
	\begin{subfigure}{.475\linewidth}
		\includegraphics[width=\linewidth]{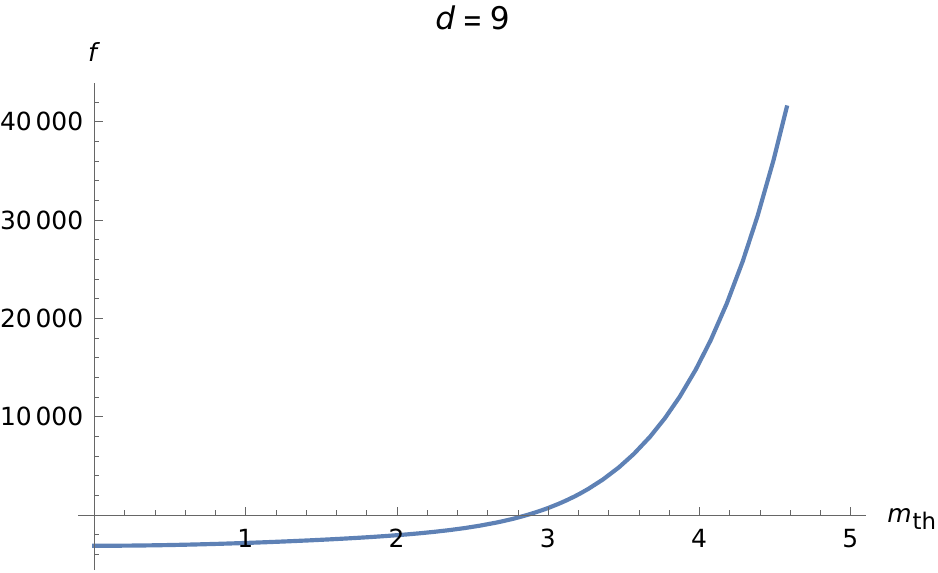}
		\caption{}
		\label{estcomp}
	\end{subfigure}
	
	\caption{The LHS of the gap equation \eqref{gap eq f} for the 
	Gross-Neveu model (denoted by $f$ in the figure) has been plotted as a function of $m_{th}$, for $d=5$ and $d=9$ the graph cuts the $x$-axis only once while for $d=3,7$, it is always positive.}
	\label{fig:rocf}
\end{figure}

We repeat the same analysis for the  Gross-Neveu model. 
The gap equation for this model at strong coupling in odd $d$ dimensions is given by,
\begin{align}\label{gap eq f}
	(m_{\rm th})^{2k}+2\sqrt{\pi}\sum _{m=0}^{k-1} \frac{ (k-m)_{2m} (2m_{\rm th}   )^{k-m} \text{Li}_{k+m}\left(-e^{-m_{\rm th}   }\right)}{m!\Gamma(\frac{1}{2}-k) }=0.
\end{align}
In contrast to the case of $O(N)$ model, the above gap equation has a positive real solution for $m_{\rm th}$ in $d=5,9,13,\ldots$ and no real solution is found for $d=3,7,11,\ldots$. This is seen in figure (\ref{fig:rocf}). 

\begin{figure}[h]
	\centering
	\includegraphics[scale=.7]{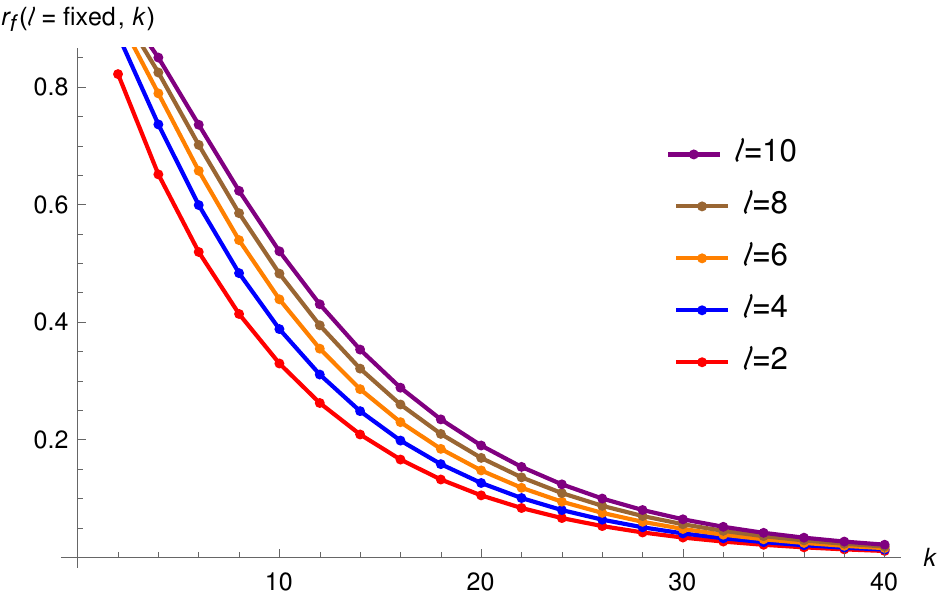}
	\caption{The ratio $ r_f(l,k) $ for the Gross-Neveu model
	 is plotted against $ k $ for various fixed values of spin $ l $. The plot shows that the one-point functions at fixed $ l $ become smaller in comparison with the same for the free theory with increasing $ d $.}
	\label{ratio fermions}
\end{figure}

\begin{figure}[h]
	\centering
	\includegraphics[scale=.7]{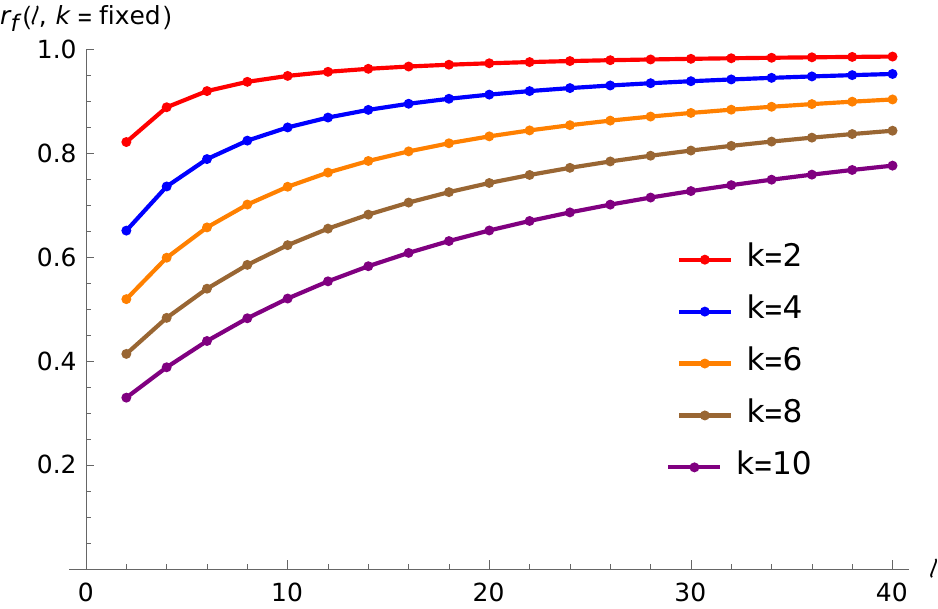}
	\caption{Plot  of  the ratio $r_f(l,k)$ for the Gross-Neveu model versus $l$, keeping $k$ fixed.}
	\label{large J fer}
\end{figure}

For the  Gross-Neveu model we have the general expression for the thermal expectation value of the operators of the kind $\mathcal{O}_+$ for arbitrary $l$ in equation \eqref{gnonep}. The ratio of one-point functions of the fermionic operators $\mathcal{O}_+$ at the non-trivial critical point to the same at the Gaussian fixed point is given by,
\begin{align}
	r_f(l,k)=\frac{a_{\mathcal{O}+}[0,l]_{m_{\rm th}, k}}{a_{\mathcal{O}+}[0,l]_{m_{\rm th}=0}}.
\end{align}
\noindent
Just as in the 
 $O(N)$ model, first we examine the behaviour of $ r_f(l,k) $ with increasing $ k $ keeping $l$ fixed. But here we restrict $ k $ to be even, as the real solution for $ m_{\rm th}  $ exists only for even $ k $. The result of this analysis is shown in figure \ref{ratio fermions}. The results are identical  to that of $O(N)$ model. 
 The ratio of the one-point function of double twist operators at fixed value of spin $ l $ evaluated at the critical point to that at the Gaussian fixed point decreases with increasing $ k $, but this ratio falls slower for higher values of $ l $.
\begin{align}
	\lim_{k\to \infty} r_f(l={\rm fixed},k)\to 0 .
\end{align}
Next, we keep the $ k $ fixed and increase $ l $ and observe how $ r(l,k) $ behaves. The result of our numerical study is described in figure \ref{large J fer}. At a fixed $k$, $r(l,k)$ tends to 1 for large values of $l$, which indicates that the one-point function of large spin operators evaluated at the critical point of Gross-Neveu model is equal to that in free theory.
\begin{align}
	\lim_{l\to\infty}r_f(l,k={\rm fixed}) =1.
\end{align}
\noindent

\begin{figure}[t]
	\centering
	\includegraphics[scale=.7]{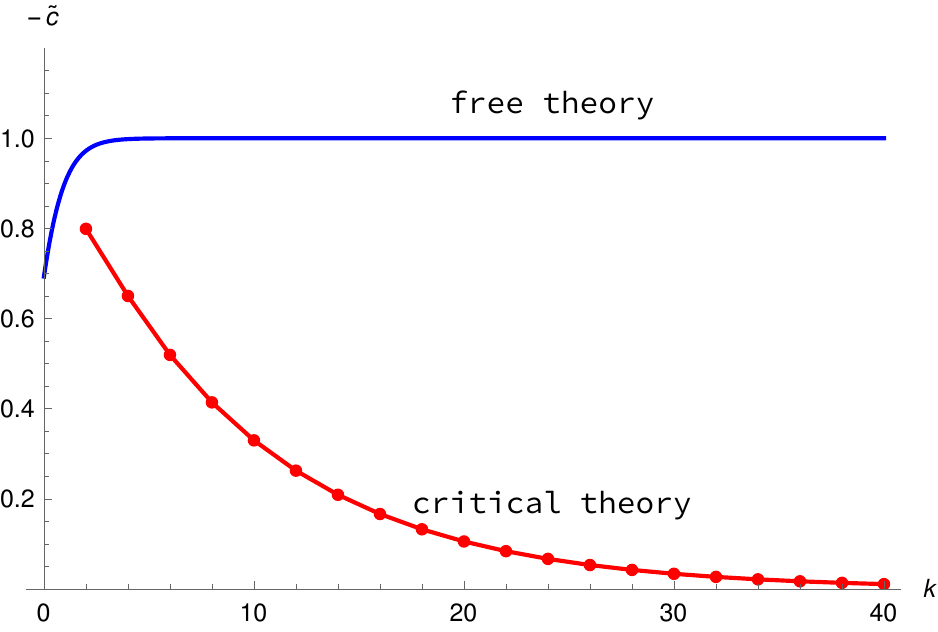}
	\caption{Plot of $\tilde c (m_{\rm th}, k ) $ vs $k$ for both the free theory and critical Gross-Neveu model.}
	\label{f fer}
\end{figure}

For a conformal field theory of fermions we can define the energy density 
\begin{equation}
T_{00} = - E =   \frac{ (d-1)  \Gamma( \frac{d}{2} )}{\pi^{\frac{d}{2}} \beta^d} \times 2^{k} \; \tilde c(m_{\rm th}, k ) .
\end{equation}
Here we factor out $2^k$ which is the number of components of the Dirac spinor in  $2k+1$ dimensions. 
We can study the behaviour of the degrees of freedom by studying $\tilde c(m_{\rm th}, k)$.
The free energy density per fermions for the  Gross-Neveu model is given by,
\begin{align} \label{energyden}
	 E =\sum _{m=0}^{k+1} \frac{ \left((k+m)^2+(k-m)\right) \Gamma (k+m) \text{Li}_{k+m}\left(-e^{-m_{\rm th}   }\right)}{m_{\rm th}^{m-k-1}\pi^k 2^m m! \Gamma (k-m+2)}-\frac{m_{th} ^{2 k+1} \Gamma \left(-k-\frac{1}{2}\right)}{2^{k+2}\pi^{k+\frac{1}{2}}}.
\end{align}
The Stefan-Boltzmann value at the Gaussian fixed point can be obtained by taking $m_{\rm th}=0$ in the above 
expression, this yields
\begin{equation}
\tilde c(m_{\rm th} =0, k) =  -( 1- 2^{-2k}) \zeta( 2k +1) .
\end{equation}
At the non-trivial fixed point, we solve the gap equation for $m_{\rm th}$ numerically  and the substitute in (\ref{energyden}) 
to obtain 
$ \tilde c(m_{\rm th}, k ) $. The result of this analysis is shown in figure \ref{f fer}. We have also plotted  the corresponding 
 behaviour of the Stefan-Boltzmann value for fermions. 
Again the effective degrees of freedom measured by the energy density decreases for the critical point as the dimension 
$d$ is increased. 
It will be interesting to understand this behaviour of one-point functions more deeply and see if this  behaviour is seen 
for all CFT's which are not free. 

\section{Conclusions} \label{secconc}

In this paper we have used the OPE inversion formula on thermal two-point functions of fermions in the 
channel which contains the stress tensor.  The OPE inversion formula was applied to the 
MFT of fermions and the large $N$  critical Gross-Neveu model. 
We studied the properties of the resulting  thermal one-point functions. 
The inversion formula made it easy to study the behaviour of the one-point functions at large 
spin as well as arbitrary dimensions. 

There are other theories where the OPE inversion formula could be applied. 
One such class of theories are the large $N$ Chern-Simons matter theories \cite{Giombi:2011kc,Aharony:2011jz}. 
These models are connected to the Gross-Neveu or the $O(N)$ vector model. 
It would be interesting to obtain thermal one-point functions in these models and study their 
spin dependence and also see how the boson-fermion duality exhibited by these models are 
reflected in the one-point functions. 
Thermal correlators in these models  are known in momentum space 
\cite{Giombi:2011kc,Jain:2012qi,Aharony:2012ns,Gur-Ari:2016xff,Ghosh:2019sqf,Mishra:2020wos}, so perhaps one way 
of proceeding in these models is to derive a OPE inversion formula directly in momentum space.

One of our motivations to study thermal one-point functions in detail was that 
one-point functions of conformal primaries evaluated in AdS black holes 
can be used to probe the interior geometry of black holes \cite{Grinberg:2020fdj,Rodriguez-Gomez:2021pfh,McInnes:2022pig,Georgiou:2022ekc,Berenstein:2022nlj,David:2022nfn}. 
The time to the singularity in the interior could be obtained as a phase factor by  suitable 
analytic continuation of the conformal dimension. 
However the one-point functions, even that of the fermionic  MFT studied here 
or  the bosonic one  in \cite{Iliesiu:2018fao}, do not exhibit this feature. 
It will be interesting to see if the application of the OPE inversion formula to the 
holographic two-point functions 
evaluated in \cite{Dodelson:2022yvn,Bhatta:2022wga,Dodelson:2023vrw} 
can be used to show that holographic one-point functions contain 
information about the interior geometry of the black hole.

In this paper we studied one-point functions in the geometry $S^1\times R^{d-1}$, 
other geometries which are relevant to holography and evaluation of entanglement entropy 
are hyperbolic cylinders or 
the  $S^1 \times AdS_{d-1}$ geometry.  
Black holes with hyperbolic horizons are dual to conformal field theories on this background. 
It should be possible to obtain an  OPE inversion formula for field theories on such backgrounds.
Recently  a proposal to write two-point functions on such curved backgrounds have been given in 
\cite{Parisini:2022wkb}.  It would be interesting to use the OPE inversion formula  to these two-point functions and 
study the properties of the resulting one-points functions in these geometries.

\appendix
\section{
Gross-Neveu model: partition function, gap equation}
\label{appendgap}

In this appendix we evaluate the partition function of the critical Gross-Neveu model at large $N$. 
We  obtain the gap equation as the saddle point equation at large $N$ and then 
evaluate its stress tensor.  
We begin with the Lagrangian  of $N$  massless  Dirac fermions $\psi^a, a= 1, 2, \cdots N$ transforming in the 
fundamental of $U(N)$  along with the 4-fermi interaction. 
\begin{equation}
S= \int d^{2k +1} x \left[ 
 \bar \psi ( i \gamma^\mu \partial_\mu ) \psi  + \frac{\lambda}{N} ( \bar \psi \psi)^2 \right].
\end{equation}
Since we are in Euclidean space $\bar \psi = \psi^\dagger$ and the $\gamma$ matrices obey
\begin{equation}
\{\gamma^\mu , \gamma^\nu \} = 2\delta^{\mu\nu}.
\end{equation}
The partition function  of the theory $S^1\times R^{2k}$ is given by 
\begin{equation}
\tilde Z= \int {\cal D}\bar\psi {\cal D} \psi e^{ - S(\psi, \bar \psi) }.
\end{equation}
We first linearise the theory using the Hubbard-Stratonovich transformation
\begin{eqnarray}
\tilde Z &=&   \int {\cal D}\bar\psi {\cal D} \psi \exp \left[ { - \int_0^\beta  d\tau d^{2k} x 
\Big(
i \bar \psi  \gamma^\mu \partial_\mu  \psi  + \frac{\lambda}{N} (\bar \psi \psi)^2 \Big) } \right], \\ \nonumber
 &=&  \int {\cal D}\bar\psi {\cal D} \psi {\cal D} \zeta
 \left[  - \int_0^\beta  d\tau d^{2k}x  \Big(
  i \bar \psi  \gamma^\mu \partial_\mu   \psi  + \frac{\zeta^2 N}{\lambda}  + i \zeta (\bar \psi \psi)
   \Big)  \right].
\end{eqnarray}
In the second line we have absorbed the normalization of the Gaussian integral over $\zeta$ into the measure. 
We can separate the zero mode of $\zeta$ and the non-zero modes and write the partition function 
\begin{eqnarray}
\zeta = \tilde \zeta + \zeta_0 .
\end{eqnarray}
Here $\zeta_0$ is the zero mode, substituting  for $\zeta$ we can write the partition function as
\begin{eqnarray}
&&\tilde Z = \int d\zeta_0  {\cal D} \bar\psi {\cal D}\psi \left[ 
\exp\Big(  -\frac{\zeta_0^2 N \beta V_{2k}}{4\lambda} 
\Big)  \exp ( -S_0 - S_I) \right] , \\ \nonumber
&&S_0 = \int d \tau d^{2k}x\Big[ i \bar \psi \gamma^\mu \partial_\mu \psi  + i \zeta_0 \bar \psi \psi  
+ \frac{\tilde \zeta^2 N}{4\lambda} \Big], \qquad 
\qquad S_I =   \int d\tau d^{2k} x \;  i \tilde\zeta \bar \psi \psi .
\end{eqnarray}
We can neglect $S_I$ in the leading large $N$ limit, observe that on canonically 
normalising the quadratic term in $\tilde \zeta$, the interaction $S_I$ acquires a factor of $\frac{1}{\sqrt{N}}$. 
After performing the Gaussian integration in $\tilde \zeta$, we are left with 
\begin{eqnarray} \label{partwithz}
\tilde Z &=&  \int d\zeta_0  \exp\left[  -\beta V_{2k} N \Big( \frac{ \zeta_0^2}{4\lambda}  - \frac{1}{\beta} \log Z ( \zeta_0)
\Big)
 \right],
\end{eqnarray}
where 
\begin{eqnarray}
 \log Z ( \zeta_0) =  2^{k-1} \sum_{n =-\infty}^\infty \int \frac{d^{2k} p}{(2\pi)^{2k}}  \log
 \left[ \frac{  4\pi ^2( n + \frac{1}{2} )^2 }{\beta^2}  + \vec p^2 + \zeta_0^2 \right].
 \end{eqnarray}
 The $2^{k-1}$ factor arises from the fact that the Dirac operator is a $2^k \times 2^k$ dimensional matrix. 
After evaluating the Matsubara sum, we obtain 
\begin{eqnarray} \label{aftermat}
\log Z(\zeta_0) =  \frac{ 2^{k-1}}{\beta^{2k}}  \int \frac{d^{2k} p}{(2\pi)^{2k}} 
\left[  \sqrt{ \vec p^2 + \zeta_0^2 \beta} +  2 \log \Big(1+ e^{ - \sqrt{ \vec p^2 + \zeta_0^2 \beta^2 } }  \big)
\right].
\end{eqnarray}
To integrate the first term in the square bracket we resort to  the  analytical continuation  of the 
integral 
\begin{equation}
\int_0^\infty  dx\; \frac{ x^{2k -1} }{ ( x^2+1)^a }  =  \frac{ \Gamma( a - k)  \Gamma( k ) }{ 2\Gamma (a) } .
\end{equation}
 The integral involving the second term in the square bracket of (\ref{aftermat}) is convergent and after some 
 straightforward manipulations can be written in terms of Polylogarithms. 
 This leads us to 
 \begin{eqnarray} \label{partitionatm}
 \log Z(\zeta_0) =-\frac{m ^{2 k+1} \beta}{\pi^k 2^{k+2}} \left[\sum _{n=0}^k \frac{   (k-n+1)_{2 n} (\zeta_0  \beta )^{-k-n-1} 
 {\rm Li}_{k+n+1}\left(-e^{-\zeta_0  \beta }\right)}{2^{n-k-2}n!}
	+\frac{  \Gamma \left(-k-\frac{1}{2}\right)}{\sqrt{\pi }}\right]. \nonumber \\
 \end{eqnarray}
 
 We can obtain the partition function $\tilde Z$ in (\ref{partwithz}) by using the saddle point approximation to perform the integral over
 $\zeta_0$. The saddle point $ \zeta_0^* = m_{\rm th} $ is determined by the equation 
 \begin{equation}
m_{\rm th}  = 2\lambda \frac{1}{\beta} \frac{\partial}{\partial m_{\rm th} } Z(m_{\rm th}) .
\end{equation}
For the critical Gross-Neveu model we take the large $\lambda$ limit.  To the leading order in the large $\lambda$ 
expansion,  the 
 saddle point equation  reduces to 
\begin{equation}
 \frac{\partial }{\partial m_{\rm th}} Z(m_{\rm th})  =0.
 \end{equation}
 This results in the following gap equation for the critical value $m_{\rm th}$. 
 \begin{align} \label{appengapeq}
2 (2m_{\rm th}\beta )^k \sum _{n=0}^{k-1} \frac{(k-n)_{2n} }{ (2 m_{\rm th}  \beta )^n n!}  \text{Li}_{k+n}(-e^{-m_{\rm th}  \beta }) +\frac{(m_{\rm th} \beta)^{2k}}{\sqrt{\pi}} \Gamma\big(\frac{1}{2}-k\big) =0.
\end{align}
 We can evaluate the stress tensor at the critical point from the partition function by 
 \begin{equation}
 T_{00} = - \frac{ \partial}{\partial \beta } \log Z( m_{\rm th}) .
 \end{equation}
 The above expression yields the energy density divided by the number of fermions $N$. 
 Performing this differentiation on the partition function given in (\ref{partitionatm}), we obtain 
 \begin{align}
	T_{00}&=-\frac{m_{\rm th} ^{2 k+1} \Gamma \left(-k-\frac{1}{2}\right)}{2^{k+2}\pi^{k+\frac{1}{2}}} \\ \nonumber
	& \qquad\qquad +
	\frac{m_{\rm th}^{k+1}}{ \pi^k \beta^k  } 
	\sum _{n=0}^{k+1} \frac{  \left[(k+n)^2+(k-n)\right] (k-n+2)_{2 n -2}}
	{(2m_{\rm th} \beta )^n n! }  \text{Li}_{k+n}\big(-e^{-m_{\rm th}  \beta }\big).
\end{align}
Note that as expected for the  critical theory, this energy density can be written as 
\begin{equation}
T_{00} = T^{d}\,  H( m_{\rm th} \beta) .
\end{equation}
where $H$ is a function of 
$m_{\rm th} \beta$,  the dimensionless number which is the root of the gap equation (\ref{appengapeq}).

\bibliographystyle{JHEP}
\bibliography{ope} 
\end{document}